\documentclass[sigconf]{acmart}

\AtBeginDocument{%
\providecommand\BibTeX{{\normalfont B\kern-0.5em{\scshape i\kern-0.25em b}\kern-0.8em\TeX}}}

\hypersetup{draft}

\usepackage[T1]{fontenc}
\usepackage{enumitem}
\usepackage{afterpage}
\usepackage{listings}
\usepackage{xcolor}
\usepackage{graphicx}
\usepackage{float}
\usepackage{caption}
\usepackage{subcaption}
\usepackage{balance}
\usepackage{placeins}
\usepackage{url}
\usepackage{epstopdf}
\usepackage{flushend}
\usepackage{booktabs}
\usepackage{footnote}
\usepackage[htt]{hyphenat}
\makesavenoteenv{table}
\makesavenoteenv{tabular}
\usepackage[acronyms,nonumberlist,nopostdot,nomain,nogroupskip]{glossaries}
\usepackage{tabularx}

\newacronym{quic}{QUIC}{Quick UDP Internet Connections}
\newacronym{3gpp}{3GPP}{3rd Generation Partnership Project}
\newacronym{adc}{ADC}{Analog to Digital Converter}
\newacronym{5g}{5G}{5th generation}
\newacronym{aimd}{AIMD}{Additive Increase Multiplicative Decrease}
\newacronym{am}{AM}{Acknowledged Mode}
\newacronym{amc}{AMC}{Adaptive Modulation and Coding}
\newacronym{aqm}{AQM}{Active Queue Management}
\newacronym{awgn}{AGWN}{Additive White Gaussian Noise}
\newacronym{balia}{BALIA}{Balanced Link Adaptation}
\newacronym{bdp}{BDP}{Bandwidth-Delay Product}
\newacronym{bf}{BF}{Beamforming}
\newacronym{cc}{CC}{Congestion Control}
\newacronym{cdf}{CDF}{Cumulative Distribution Function}
\newacronym{cn}{CN}{Core Network}
\newacronym{cqi}{CQI}{Channel Quality Information}
\newacronym{cp}{CP}{Control Plane}
\newacronym{csirs}{CSI-RS}{Channel State Information - Reference Signal}
\newacronym{dc}{DC}{Dual Connectivity}
\newacronym{dce}{DCE}{Direct Code Execution}
\newacronym{dci}{DCI}{Downlink Control Information}
\newacronym{dl}{DL}{Downlink}
\newacronym{dmr}{DMR}{Deadline Miss Ratio}
\newacronym{dmrs}{DMRS}{DeModulation Reference Signal}
\newacronym{e2e}{E2E}{End-to-End}
\newacronym{ecn}{ECN}{Explicit Congestion Notification}
\newacronym{edf}{EDF}{Earliest Deadline First}
\newacronym{enb}{eNB}{evolved Node Base}
\newacronym{epc}{EPC}{Evolved Packet Core}
\newacronym{es}{ES}{Edge Server}
\newacronym{fdma}{FDMA}{Frequency Division Multiple Access}
\newacronym{fdd}{FDD}{Frequency Division Duplexing}
\newacronym[firstplural=Radio Access Technologies (RATs)]{rat}{RAT}{Radio Access Technology}
\newacronym{fs}{FS}{Fast Switching}
\newacronym{ftp}{FTP}{File Transfer Protocol}
\newacronym{gnb}{gNB}{Next Generation Node Base}
\newacronym{harq}{HARQ}{Hybrid Automatic Repeat reQuest}
\newacronym{hetnet}{HetNet}{Heterogeneous Network}
\newacronym{hh}{HH}{Hard Handover}
\newacronym{hol}{HOL}{Head-of-Line}
\newacronym{ia}{IA}{Initial Access}
\newacronym{imt}{IMT}{International Mobile Telecommunication}
\newacronym{iot}{IoT}{Internet of Things}
\newacronym{los}{LOS}{Line of Sight}
\newacronym{lte}{LTE}{Long Term Evolution}
\newacronym{m2m}{M2M}{Machine to Machine}
\newacronym{mac}{MAC}{Medium Access Control}
\newacronym{mc}{MC}{Multi-Connectivity}
\newacronym{mcs}{MCS}{Modulation and Coding Scheme}
\newacronym{mec}{MEC}{Mobile Edge Cloud}
\newacronym{mi}{MI}{Mutual Information}
\newacronym{mimo}{MIMO}{Multiple-Input Multiple-Output}
\newacronym{mmwave}{mmWave}{millimeter wave}
\newacronym{mr}{MR}{Maximum Rate}
\newacronym{mss}{MSS}{Maximum Segment Size}
\newacronym{mtd}{MTD}{Machine-Type Device}
\newacronym{mtu}{MTU}{Maximum Transmission Unit}
\newacronym{nfv}{NFV}{Network Function Virtualization}
\newacronym{nlos}{NLOS}{Non Line of Sight}
\newacronym{nr}{NR}{New Radio}
\newacronym{ofdm}{OFDM}{Orthogonal Frequency Division Multiplexing}
\newacronym{pdcch}{PDCCH}{Physical Downlonk Control Channel}
\newacronym{pdcp}{PDCP}{Packet Data Convergence Protocol}
\newacronym{pdsch}{PDSCH}{Physical Downlink Shared Channel}
\newacronym{pdu}{PDU}{Packet Data Unit}
\newacronym{pf}{PF}{Proportional Fair}
\newacronym{pgw}{PGW}{Packet Gateway}
\newacronym{phy}{PHY}{Physical}
\newacronym{pbch}{PBCH}{Physical Broadcast Channel}
\newacronym[plural=\gls{mme}s,firstplural=Mobility Management Entities (MMEs)]{mme}{MME}{Mobility Management Entity}
\newacronym{prb}{PRB}{Physical Resource Block}
\newacronym{pss}{PSS}{Primary Synchronization Signal}
\newacronym{pucch}{PUCCH}{Physical Uplink Control Channel}
\newacronym{pusch}{PUSCH}{Physical Uplink Shared Channel}
\newacronym{rach}{RACH}{Random Access Channel}
\newacronym{ran}{RAN}{Radio Access Network}
\newacronym{red}{RED}{Random Early Detection}
\newacronym{rf}{RF}{Radio Frequency}
\newacronym{rlc}{RLC}{Radio Link Control}
\newacronym{rlf}{RLF}{Radio Link Failure}
\newacronym{rrc}{RRC}{Radio Resource Control}
\newacronym{rrm}{RRM}{Radio Resource Management}
\newacronym{rr}{RR}{Round Robin}
\newacronym{rs}{RS}{Remote Server}
\newacronym{rsrp}{RSRP}{Reference Signal Received Power}
\newacronym{rss}{RSS}{Received Signal Strength}
\newacronym{rtt}{RTT}{Round Trip Time}
\newacronym{rw}{RW}{Receive Window}
\newacronym{rx}{RX}{Receiver}
\newacronym{sa}{SA}{standalone}
\newacronym{sack}{SACK}{Selective Acknowledgment}
\newacronym{sap}{SAP}{Service Access Point}
\newacronym{sch}{SCH}{Secondary Cell Handover}
\newacronym{scoot}{SCOOT}{Split Cycle Offset Optimization Technique}
\newacronym{sdma}{SDMA}{Spatial Division Multiple Access}
\newacronym{sinr}{SINR}{Signal to Interference plus Noise Ratio}
\newacronym{sm}{SM}{Saturation Mode}
\newacronym{snr}{SNR}{Signal to Noise Ratio}
\newacronym{son}{SON}{Self-Organizing Network}
\newacronym{ss}{SS}{Synchronization Signal}
\newacronym{srs}{SRS}{Sounding Reference Signal}
\newacronym{sss}{SSS}{Secondary Synchronization Signal}
\newacronym{tb}{TB}{Transport Block}
\newacronym{tcp}{TCP}{Transmission Control Protocol}
\newacronym{tdd}{TDD}{Time Division Duplexing}
\newacronym{tdma}{TDMA}{Time Division Multiple Access}
\newacronym{tfl}{TfL}{Transport for London}
\newacronym{tm}{TM}{Transparent Mode}
\newacronym{trp}{TRP}{Transmitter Receiver Pair}
\newacronym{tti}{TTI}{Transmission Time Interval}
\newacronym{ttt}{TTT}{Time-to-Trigger}
\newacronym{tx}{TX}{Transmitter}
\newacronym{ue}{UE}{User Equipment}
\newacronym{ul}{UL}{Uplink}
\newacronym{uml}{UML}{Unified Modeling Language}
\newacronym{um}{UM}{Unacknowledged Mode}
\newacronym{utc}{UTC}{Urban Traffic Control}
\newacronym{vm}{VM}{Virtual Machine}
\newacronym{rsrq}{RSRQ}{Reference Signal Received Quality}
\newacronym{rssi}{RSSI}{Received Signal Strength Indicator}
\newacronym{crs}{CRS}{Cell Reference Signal}
\newacronym{comp}{CoMP}{Coordinated Multi-Point}
\newacronym{cran}{C-RAN}{Cloud \acrlong{ran}}
\newacronym{ca}{CA}{Carrier Aggregation}
\newacronym{cco}{CC}{Carrier Component}
\newacronym{nsa}{NSA}{Non Stand Alone}
\newacronym{embb}{eMBB}{Enhanced Mobility Broadband}
\newacronym{bsr}{BSR}{Buffer Status Report}
\newacronym{srb}{SRB}{Service Radio Bearer}
\newacronym{scm}{SCM}{Spatial Channel Model}
\newacronym{sctp}{SCTP}{Stream Control Transmission Protocol}
\newacronym{mptcp}{MPTCP}{Multi-path TCP}
\newacronym{ietf}{IETF}{Internet Engineering Task Force}
\newacronym{os}{OS}{Operating System}
\newacronym{tls}{TLS}{Transport Layer Security}
\newacronym{rfc}{RFC}{Request for Comments}
\newacronym{http}{HTTP}{HyperText Transfer Protocol}
\newacronym{nat}{NAT}{Network Address Translation}
\newacronym{api}{API}{Application Programming Interface}
\newacronym{rto}{RTO}{Retransmission Timeout}
\newacronym{psc}{PSC}{Public Safety Communication}
\newacronym{rpgm}{RPGM}{Reference Point Group Mobility}
\newacronym{ic}{IC}{Incident Command}
\newacronym{rsu}{RSU}{Road Side Unit}
\newacronym{uav}{UAV}{Unmanned Aerial Vehicle}
\newacronym{iab}{IAB}{Integrated Access and Backhaul}
\newacronym{psd}{PSD}{Power Spectral Density}
\newacronym{mpc}{MPC}{Multi Path Component}
\newacronym{rt}{RT}{Ray Tracer}
\newacronym{aoa}{AoA}{Angle of Arrival}
\newacronym{aod}{AoD}{Angle of Departure}
\newacronym{inr}{INR}{Interference to Noise Ratio}
\newacronym{qd}{QD}{Quasi Deterministic}
\newacronym{wlan}{WLAN}{Wireless Local Area Network}
\newacronym{cad}{CAD}{Computer-aided Design}
\newacronym{ap}{AP}{Access Point}
\newacronym{sta}{STA}{Station}
\newacronym{nrmse}{NRMSE}{Normalized Root Mean Square Error}
\newacronym{ut}{UT}{User Terminal}
\newacronym{bs}{BS}{Base Station}

\usepackage{tikz}
\usepackage{pgfplots}
\pgfplotsset{compat=newest} 
\pgfplotsset{plot coordinates/math parser=false} 
\newlength\fheight
\newlength\fwidth
\usetikzlibrary{plotmarks,patterns,decorations.pathreplacing,backgrounds,calc,arrows,arrows.meta,spy,matrix}
\usepgfplotslibrary{patchplots,groupplots}
\usepackage{tikzscale}

\tikzstyle{startstop} = [rectangle, rounded corners, minimum width=2cm, minimum height=0.5cm,text centered, draw=black]
\tikzstyle{io} = [trapezium, trapezium left angle=70, trapezium right angle=110, minimum width=3cm, minimum height=1cm, text centered, draw=black]
\tikzstyle{process} = [rectangle, minimum width=2cm, minimum height=0.5cm, text centered, draw=black, align=center]
\tikzstyle{decision} = [ellipse, minimum width=2cm, minimum height=1cm, text centered, draw=black]
\tikzstyle{arrow} = [thick,<->,>=stealth]
\tikzstyle{line} = [thick,>=stealth]
\tikzstyle{darrow} = [thick,<->,>=stealth,dashed]
\tikzstyle{sarrow} = [thick,->,>=stealth]
\tikzstyle{larrow} = [line width=0.1mm,dashdotted,<->,>=stealth]

\definecolor{SchoolColor}{RGB}{0.71, 0, 0.106}
\definecolor{chaptergrey}{rgb}{0.61, 0, 0.09} 
\definecolor{midgrey}{rgb}{0.4, 0.4, 0.4}
\definecolor{chaptergreen}{rgb}{0.09, 0.612, 0}
\definecolor{chapterpurple}{rgb}{0.522, 0, 0.612}
\definecolor{chapterlightgreen}{rgb}{0, 0.612, 0.522}

\makeatletter
\def\grd@save@target#1{%
  \def\grd@target{#1}}
\def\grd@save@start#1{%
  \def\grd@start{#1}}
\tikzset{
  grid with coordinates/.style={
    to path={%
      \pgfextra{%
        \edef\grd@@target{(\tikztotarget)}%
        \tikz@scan@one@point\grd@save@target\grd@@target\relax
        \edef\grd@@start{(\tikztostart)}%
        \tikz@scan@one@point\grd@save@start\grd@@start\relax
        \draw[minor help lines] (\tikztostart) grid (\tikztotarget);
        \draw[major help lines] (\tikztostart) grid (\tikztotarget);
        \grd@start
        \pgfmathsetmacro{\grd@xa}{\the\pgf@x/1cm}
        \pgfmathsetmacro{\grd@ya}{\the\pgf@y/1cm}
        \grd@target
        \pgfmathsetmacro{\grd@xb}{\the\pgf@x/1cm}
        \pgfmathsetmacro{\grd@yb}{\the\pgf@y/1cm}
        \pgfmathsetmacro{\grd@xc}{\grd@xa + \pgfkeysvalueof{/tikz/grid with coordinates/major step x}}
        \pgfmathsetmacro{\grd@yc}{\grd@ya + \pgfkeysvalueof{/tikz/grid with coordinates/major step y}}
        \foreach \x in {\grd@xa,\grd@xc,...,\grd@xb}
        \node[anchor=north] at (\x,\grd@ya) {\pgfmathprintnumber{\x}};
        \foreach \y in {\grd@ya,\grd@yc,...,\grd@yb}
        \node[anchor=east] at (\grd@xa,\y) {\pgfmathprintnumber{\y}};
      }
    }
  },
  minor help lines/.style={
    help lines,
    gray,
    line cap =round,
    xstep=\pgfkeysvalueof{/tikz/grid with coordinates/minor step x},
    ystep=\pgfkeysvalueof{/tikz/grid with coordinates/minor step y}
  },
  major help lines/.style={
    help lines,
    line cap =round,
    line width=\pgfkeysvalueof{/tikz/grid with coordinates/major line width},
    xstep=\pgfkeysvalueof{/tikz/grid with coordinates/major step x},
    ystep=\pgfkeysvalueof{/tikz/grid with coordinates/major step y}
  },
  grid with coordinates/.cd,
  minor step x/.initial=.5,
  minor step y/.initial=.2,
  major step x/.initial=1,
  major step y/.initial=1,
  major line width/.initial=1pt,
}
\makeatother

\begin{document}

\title{Implementation of a Spatial Channel Model for ns-3}

\author{Tommaso Zugno}
\author{Michele Polese}
\affiliation{%
  \institution{Department of Information Engineering,\\University of Padova}
  \city{Padova}
  \state{Italy}
}
\email{{zugnotom,polesemi}@dei.unipd.it}

\author{Natale Patriciello}
\author{Biljana Bojovi\'c}
\author{Sandra Lagen}
\affiliation{%
  \institution{Centre Tecnol\`ogic de Telecomunicacions de Catalunya (CTTC/CERCA)}
  \streetaddress{Av. C.F. Gauss, 7}
  \city{Castelldefels, Barcelona}
  \state{Spain}
  \postcode{08013}
}
\email{{npatriciello, bbojovic, slagen}@cttc.es}

\author{Michele Zorzi}
\affiliation{%
  \institution{Department of Information Engineering,\\University of Padova}
  \city{Padova}
  \state{Italy}
}
\email{zorzi@dei.unipd.it}

\copyrightyear{2020}
\acmYear{2020}
\setcopyright{acmcopyright}
\acmConference[WNS3 2020]{2020 Workshop on ns-3}{June 17--18, 2020}{Gaithersburg, MD, USA}
\acmBooktitle{2020 Workshop on ns-3 (WNS3 2020), June 17--18, 2020, Gaithersburg, MD, USA}
\acmPrice{15.00}
\acmDOI{10.1145/3389400.3389401}
\acmISBN{978-1-4503-7537-5/20/06}

\pagestyle{empty}

\begin{abstract}

The next generation of wireless networks will feature a more flexible radio access design, integrating multiple new technological solutions (e.g., massive \gls{mimo}, millimeter waves) to satisfy different verticals and use cases. The performance evaluation of these networks will require more complex models to represent the interactions of different components of the networks accurately. For example, channel models, which are of paramount importance to precisely characterize the behavior of such systems, need to account for multi-antenna systems and new frequency bands. This paper presents the ns-3 implementation of a spatial channel model for the 0.5-100 GHz spectrum, following the 3GPP Technical Report 38.901. The code, designed to be flexible and easily extensible, is integrated in ns-3's \texttt{antenna}, \texttt{propagation} and \texttt{spectrum} models, and offers the support for the investigation of future wireless systems in ns-3.

\begin{picture}(0,0)(0,-360)
\put(0,0){
\put(0,0){\footnotesize This paper has been accepted for presentation at the 2020 Workshop on ns-3 (WNS3 2020), June 17--18, 2020, Gaithersburg, MD, USA}}
\end{picture}

\end{abstract}


\maketitle

\vspace{-0.2cm}
\section{Introduction}\label{sec:intro}
\glsresetall

Wireless networks are rapidly evolving to meet the datarate, latency and reliability demands of an increasingly connected society. The \gls{5g} of cellular networks is being deployed, following the specifications of \gls{3gpp} \gls{nr}~\cite{38300}. Similarly, multiple generations of \glspl{wlan} are being proposed to achieve higher capacity, improved spatial reuse, or lower energy consumption with IEEE 802.11ax/ay/ah/bd~\cite{zhou2018ieee,khorov2019tutorial}. These wireless networking standards will be the first to exploit new technologies such as, for example, massive \gls{mimo} and \gls{mmwave} communications~\cite{BocHLMP:14}.

Standardization bodies (i.e., the \gls{3gpp} and the IEEE) are aiming at addressing the requirements of multiple verticals and use cases with a single design of the \gls{ran}~\cite{3GPP38913}. For example, \gls{3gpp} NR can be configured to support enhanced mobile broadband (to provide ultra-high capacity to the end users), ultra-reliable and low latency communications (e.g., for remote control scenarios), and massive machine-type deployments. Therefore, future wireless networks will exhibit an increasing degree of complexity and flexibility. The different versions of IEEE \glspl{wlan} and \gls{3gpp} NR can (i) operate on a wide portion of spectrum, which includes the traditional bands below 6 GHz and the \gls{mmwave} frequencies; (ii) be deployed on devices equipped with quasi-omnidirectional antennas, or with phased antenna arrays to perform analog, hybrid or digital beamforming; and (iii) support devices that communicate while moving at different speeds, up to 500 km/h for NR.

A correct and reliable testing and performance evaluation of such complex networks becomes of paramount importance to identify the critical elements of the system before commercializing it, and to understand which algorithms and network architectures can provide the best quality of service to the end users.
Simulation will play a fundamental role in this, as testbeds for 5G and next-generation \glspl{wlan} are still in the making~\cite{polese2019millimetera,saha2019x60}.
Additionally, simulations can adapt better than testbeds to the large number of evolving use cases and deployment scenarios that such networks will serve. ns-3 is well positioned to be an important simulation tool for future wireless networks, thanks to the already available modules for \glspl{mmwave} and NR~\cite{mezzavilla2017end,PATRICIELLO2019101933}, IEEE 802.11ad/ay~\cite{assasa2017extending,assasa11ad2019}, and to the activity to extend the \texttt{wifi} module to also support IEEE 802.11ax~\cite{lanante2019improved}.

Nonetheless, ns-3 is currently lacking common channel model \glspl{api} that can be used by all the aforementioned modules, to provide results based on the same channel abstraction, or to test the coexistence of different technologies in the same frequency spectrum. These modules, indeed, currently use different channel modeling techniques, included in the modules themselves~\cite{assasa2019high,rebato2018multi,zhang2017ns3}, not directly comparable with each other, and not designed with a modular and extensible approach. ns-3, on the other hand, provides a number of propagation models, and a flexible abstraction for the spectrum usage of single and multi carrier systems~\cite{baldo2009spectrum}, but is lacking a fading model that can be integrated with multi-antenna wireless technologies.

The channel model, however, is one of the most important components of a wireless network simulator, as the results can only be as accurate as the channel model~\cite{ferrand2016trends}. In particular, when it comes to \glspl{mmwave}, the harsh propagation conditions may severely impact the performance of the higher layers of the protocol stack, much more so than at traditional sub-6 GHz frequencies~\cite{polese2018impact}. Moreover, \gls{mmwave} systems generally exploit beamforming to increase the link budget of the communication, and this element has to be introduced in the overall modeling process of the channel. Additionally, when considering \gls{mimo} systems, an exact characterization of the rank of the wireless channel is necessary for a proper evaluation of how many parallel streams can be supported~\cite{spencer2004introduction}.

In this paper, we present the implementation of a channel model for future wireless networks that has been recently included in ns-3. Notably, we implemented a \gls{scm} for the \texttt{spectrum} module, which characterizes the channel through a matrix $\mathbf{H}$, in which each single entry models the channel between two antenna elements at the transmitter and the receiver~\cite{saleh1987statistical}. The channel realization is computed using the \gls{3gpp} stochastic model for 5G networks between 0.5 and 100 GHz~\cite{TR38901}. Additionally, we extended the \texttt{propagation} module to support the models in~\cite{TR38901}, with a different characterization for \gls{los} and \gls{nlos} states (according to whether the direct path between the transmitter and the receiver is blocked or not), and the \texttt{antenna} module, which now features antenna arrays. The implementation of the channel model equations is based on that in~\cite{zhang2017ns3}, but the code has been refactored and redesigned to be as modular as possible, with a clear separation of the propagation model, the fading, the antenna, and the beamforming. Moreover, it can be easily extended to support other fading models based on the computation of a channel matrix. We believe that this model, which has been developed as part of the Google Summer of Code project, represents a substantial and timely contribution to the wireless research community that uses ns-3 to study next-generation wireless networks.

The rest of the paper is organized as follows. In Section~\ref{sec:channel} we review modern channel modeling efforts, with a focus on the 3GPP channel model for 5G and on why \gls{scm}s are widely used in this context. We describe the implementation of the \gls{3gpp} \gls{scm} from~\cite{TR38901} in Section~\ref{sec:implementation}, and present examples and comment on use cases in Section~\ref{sec:scenarios}. Finally, we conclude the paper in Section~\ref{sec:conclusions}.

\vspace{-0.33cm}
\section{Recent Developments in Channel Modeling}
\label{sec:channel}
\begin{figure*}[t!]
  \setlength\abovecaptionskip{-0.03cm}
  \setlength\belowcaptionskip{-0.36cm}
  \centering
  \includegraphics[trim={0 3cm 0 4cm}, width=0.8\textwidth]{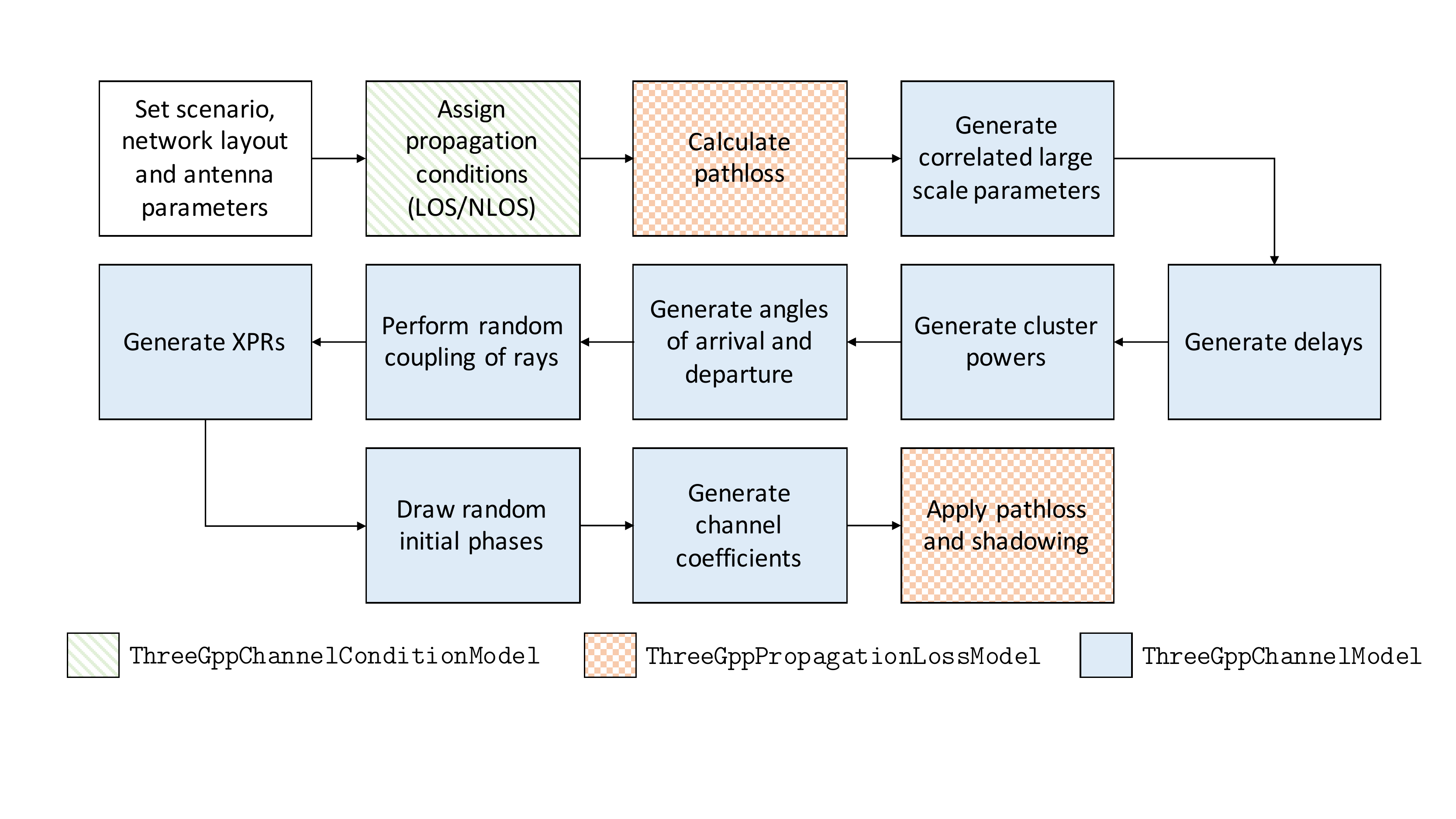}
  \caption{Diagram Representing the Channel Generation Procedure; the Colors Indicate the Classes where the Steps are Accomplished}
  \label{fig:procedure}
\end{figure*}

Channel modeling is a fundamental activity for the design and evaluation of future wireless networks. The authors of~\cite{ferrand2016trends} claim that the new features of cellular and \gls{wlan} networks call for new approaches in channel modeling. Large antenna arrays and the deployment of \gls{mimo} techniques require the addition of the spatial dimension in the channel, with a full 3D model, capable of characterizing the diversity of the channel paths for each pair of antenna elements between the transmitter and the receiver. Moreover, the channel in the new frequency bands of \gls{3gpp} NR and IEEE 802.11ad/ay (i.e., \glspl{mmwave}) needs proper understanding, especially with respect to multipath fading and blockage. Finally, new deployments (e.g., vehicular networks) introduce additional modeling requirements for network simulations.

These challenges have motivated several efforts in channel modeling, especially when considering \gls{mmwave} frequencies~\cite{hemadeh2018millimeter}. Multiple measurement campaigns in these frequency bands have strived to accurately model the propagation and fading in different scenarios~\cite{rappaportmillimeter,koymen2015indoor,remley2017measurement}, highlighting how \glspl{mmwave} are characterized by high propagation loss, sensitivity to blockage, and a reduced impact of small scale fading with sparsity in the angular domain.
These measurement campaigns have then led to different families of channel models for future wireless networks, generally given by the combination of propagation loss and fading models. The different modeling approaches differ for their degree of abstraction, simplicity and accuracy. Analytical studies for 5G generally use simple propagation loss models, combined with Nakagami-m or Rayleigh fading~\cite{andrews2017modeling}. These models are computationally efficient, but fail to capture the spatial dimension of the channel and cannot be combined with realistic beamforming models. Quasi-deterministic channels, developed, for example, for IEEE 802.11ad/ay~\cite{maltsev2016channel}, are instead designed to be as accurate as possible in specific scenarios, but are much more complex and require a precise characterization of the environment~\cite{lecci2020simplified}.

\vspace{-0.51cm}
\subsection{3GPP TR 38.901}
For the evaluation of NR, the \gls{3gpp} has adopted a 3D \gls{scm}~\cite{TR38901}, which represents a tradeoff between the two aforementioned channel modeling approaches: it is generic, thanks to its stochastic nature, but at the same time can model interactions with beamforming vectors. An \gls{scm}, indeed, models the channel through a channel matrix $\mathbf{H}(t,\tau)$, with as many rows and columns as the number of transmit ($U$) and receive ($S$) antenna elements. Each entry $H_{u,s}(t,\tau)$ corresponds to the impulse response of the channel between the $s$-th element of the \gls{bs} antenna and the $u$-th element of the \gls{ut} antenna at delay $\tau$ at time $t$.
$H_{u,s}(t,\tau)$ is generated by the superposition of $N$ different clusters, representing groups of multipath components that arrive and/or depart the antenna arrays with certain angles. The multipath components impact the receiving array with different delays, and the power will be scaled according to a delay-based profile. If present, an \gls{los} cluster is modeled with the strongest power and the minimum delay. The other clusters, instead, represent reflections from the scattering environment.

The \gls{3gpp} channel modeling framework is described in TR~38.901 \cite{TR38901} and represents the extension of TR~38.900, which was targeted for above-$6$~GHz bands only. It supports the modeling of wireless channels between $0.5$ and $100$~GHz by means of a stochastic \gls{scm}, in which a single instance of the channel matrix $\mathbf{H}(t,\tau)$ is computed according to random distributions for large scale fading parameters (i.e., the delay profile, the angles of arrival and departure, and the shadowing) and for the small scale fading (i.e., for small variations in the channel, for example, as given by the Doppler spread).
To enable the simulation of signal propagation in different environments, it specifies four scenarios, with different parameters for the random distributions underlying the channel:
\begin{itemize}
  \item RMa (Rural Macro), targeting rural deployments with continuous wide area coverage;
  \item UMa (Urban Macro), intended to model urban areas with macrocells mounted above the rooftops of the surrounding buildings;
  \item UMi (Urban Micro) Street Canyon, similar to UMa but with base stations mounted below the rooftops;
  \item Indoor Hotspot (InH) Mixed and Open Office, to model indoor environments.
\end{itemize}

For each scenario, this model provides the characterization of the \gls{los}/\gls{nlos} channel condition, the propagation loss, and the small scale fading due to the effect of Doppler and multipath. 
Also, it defines a radiation model to account for the non-isotropic behavior of real antennas.


The channel matrix generation procedure, represented in Figure \ref{fig:procedure}, accounts for both large (i.e., pathloss and shadowing) and small scale (fast fading) propagation phenomena, and provides the possibility to select different models and parameters depending on the scenario of interest.
The pathloss model describes the signal attenuation between the transmitter and the receiver as a function of the 3D positions and the carrier frequency. The shadowing model provides the statistical characterization of the attenuation due to the presence of obstacles between the transmitter and the receiver. 
The small scale fading accounts for the signal phase and amplitude variations due to small changes in the spatial separation between the transmitter and the receiver, and for the Doppler effect introduced by a moving terminal. While the large scale propagation effects are considered to be constant within the frequency band of interest, the small scale fading has a frequency-selective behavior, thus introducing a gain which varies within the band.

In the following, we describe the \gls{3gpp} \gls{scm} for 5G networks that has been implemented in ns-3, providing details on the pathloss and channel condition computations, the channel matrix generation procedure, and the antenna model that can be associated to such matrix.
\begin{figure*}[t!]
  \setlength\belowcaptionskip{-0.33cm}
  \centering
  \includegraphics[trim={0cm, 0cm, 0cm, 1cm}, width=\textwidth]{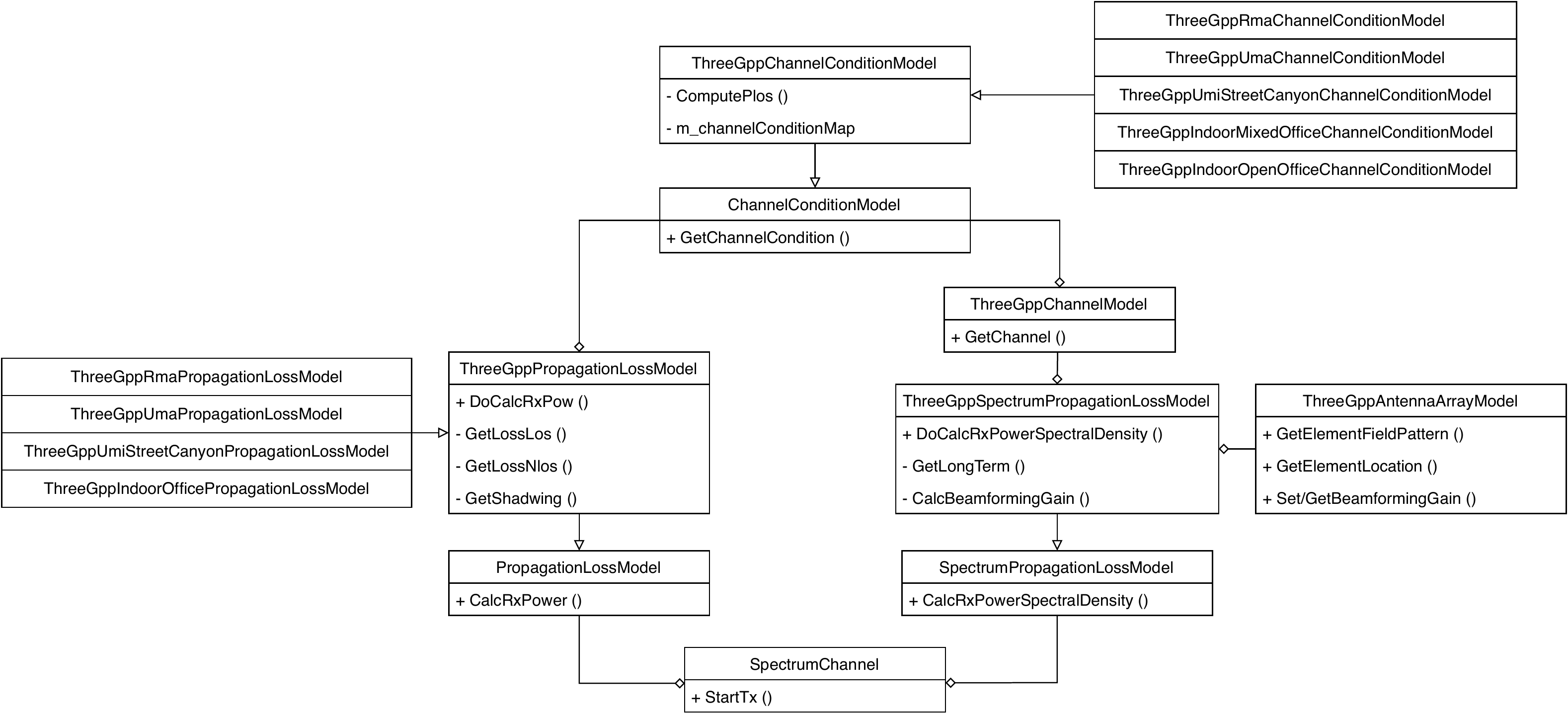}
  \caption{Simplified \gls{uml} Diagram of the \gls{scm} Implementation}
  \label{fig:uml}
\end{figure*}

\vspace{-0.2cm}
\section{ns-3 Implementation}
\label{sec:implementation}

In ns-3, the modeling of the signal propagation through the wireless channel is handled by the \texttt{spectrum} module, which includes the abstract classes \texttt{SpectrumPhy} and \texttt{SpectrumChannel}.
Devices communicating through the same wireless channel have their own \texttt{SpectrumPhy} instances, which are in charge of creating the \gls{psd} of the transmitted signals. The different \texttt{SpectrumPhy} instances are attached to the same \texttt{SpectrumChannel} object which dispatches the transmissions among the devices.
At each transmission, \texttt{SpectrumPhy} calls the method \texttt{SpectrumChannel::StartTx} which notifies each receiver and computes the corresponding \glspl{psd} of the received signals.
To account for the power attenuation and fading due to the propagation of the signal through the environment, \texttt{SpectrumChannel} relies on two standard interfaces, i.e., \texttt{PropagationLossModel} and \texttt{SpectrumPropagationLossModel}. The former models slow fading, in which the loss is constant over the frequency band of the signal, while the latter is used for fast fading models, which introduce frequency-selective losses.

The 3GPP \gls{scm} can be divided into four main components, namely, (i) channel condition models, used to determine the \gls{los}/\gls{nlos} channel state, (ii) propagation loss models, including pathloss and shadowing, (iii) the fast fading model, and (iv) the antenna model.
The objective of this project was to implement these components using, whenever possible, the interfaces provided by the \texttt{spectrum} and \texttt{propagation} modules~\cite{baldo2009spectrum}, without compromising the support of existing models and ensuring an easy integration in the main code base.
We decided to implement each component as a separate class in order to achieve a flexible and re-usable architecture, enabling the possibility to easily replace, modify or include new parts.
Figure~\ref{fig:uml} reports a simplified \gls{uml} diagram for the classes involved in the channel model implementation.

\vspace{-0.33cm}
\subsection{LOS Probability Models}
The first step for the generation of the channel matrix is to determine the LOS/NLOS channel condition. 3GPP TR~38.901 provides stochastic models to determine the channel state in all the scenarios of interest, taking into account the distance between the communication endpoints and the characteristics of the propagation environment, e.g., the presence of buildings and obstacles.

Since ns-3 lacks a general way to account for the channel state, we developed the class \texttt{ChannelCondition}, which stores the state information related to a certain channel. Also, we proposed a new interface, called \texttt{ChannelConditionModel}, which can be extended to implement any specific channel condition model, either stochastic or deterministic.
The main method is \texttt{GetChannelCondition}, which accepts as argument the positions of the two nodes and returns a pointer to the corresponding \texttt{ChannelCondition} instance.

To include the channel condition models defined in the 3GPP TR 38.901, we developed five different classes, i.e., \texttt{ThreeGppRma\-ChannelConditionModel}, \texttt{ThreeGppUmaChannelConditionModel}, \texttt{ThreeGppUmiStreetCanyonChannelConditionModel}, \texttt{ThreeGpp\-Indoor\-OpenOfficeChannelConditionModel} and \texttt{ThreeGpp\-Indoor\-MixedOfficeChannelConditionModel}, each handling a different scenario. All the new classes derive from the same base, called \texttt{ThreeGppChannelConditionModel}, which extends the \texttt{ChannelConditionModel} interface and provides caching functionalities for the periodic update of the states.
When the method \texttt{GetChannelCondition} is called for the first time, the channel state is computed and its value is stored in a map, together with the generation time.
Then, at subsequent calls, the method checks if the state has to be updated or not based on the time expired since its generation and, if so, a new state is independently generated, without accounting for any temporal correlation. The update interval can be tuned by the user with the attribute \texttt{UpdatePeriod}, with the possibility of never updating the channel condition if the attribute is set to $0$.

\vspace{-0.2cm}
\subsection{Pathloss and Shadowing Models}
The pathloss models defined in 3GPP TR~38.901 can be expressed through the general form of Eq.~\eqref{eq:pathloss}, where $d$ is the 3D distance between the two endpoints, $f_C$ is the carrier frequency, $A$, $B$ and $C$ are model parameters, and $X$ is an optional loss term.
\begin{equation}\label{eq:pathloss}
  PL = A \log_{10} (d) + B + C \log_{10} (f_C) + X \quad \text{[dB]}
\end{equation}
In particular, $A$ represents the pathloss exponent and accounts for the dependence on the distance between the receiver and the transmitter, while $C$ determines the relation between the pathloss and the carrier frequency.
$A$, $B$, $C$ and $X$ take different values depending on the propagation conditions, such as the scenario, the LOS/NLOS channel state and the break point distance $d_{BP}$, as defined in \cite{TR38901}.

Also, to account for the variations of the received signal power due to  blockage events, a log-normal shadowing component is added to the mean pathloss. Adjacent fading values are correlated with an exponential autocorrelation function, and their correlation depends on the spatial separation between the two positions. As for the pathloss, the standard deviation of the shadowing component, as well as the autocorrelation function, depend on the specific propagation conditions.

Moreover, 3GPP TR~38.901 specifies a model to account for the outdoor-to-indoor penetration loss due to buildings or cars, which however was not considered in this work and is planned for future development.

To include the pathloss and shadowing model defined in 3GPP TR~38.901, we developed the base class \texttt{ThreeGppPropagationLossModel}, which extends the \texttt{PropagationLossModel} interface and implements the general logic used to handle the computation of the mean pathloss and the shadowing component. Then, we extended this class by developing four subclasses, i.e.,  \texttt{ThreeGppRmaPropagationLossModel}, \texttt{ThreeGppUmaPropagationLossModel}, \texttt{ThreeGppUmaStreetCanyonPropagationLossModel} and\\\texttt{ThreeGppIndoorOfficePropagationLossModel}, which define the models for the different channel scenarios.
Since the propagation loss depends on the \gls{los}/\gls{nlos} channel state, the \texttt{ThreeGppPropagationLossModel} class is paired with a channel condition model through the \texttt{ChannelConditionModel} interface.
The main method is \texttt{DoCalcRxPower}, which returns the power received at the receiver side based on the positions of the communicating nodes.
It makes use of the methods \texttt{GetLossLos} and \texttt{GetLossNlos} to compute the mean pathloss in the \gls{los} and \gls{nlos} states, respectively, and of the method \texttt{GetShadowing} to apply the shadowing model.
Two other functions, namely \texttt{GetShadowingStd} and \texttt{GetShadowingCorrelationDistance}, are used by \texttt{GetShadowing} to retrieve the standard deviation of the shadowing component and the correlation distance, a parameter which defines the autocorrelation function.

\vspace{-0.2cm}
\subsection{Fast Fading Model}
The fast fading model included in 3GPP TR~38.901 accounts for the changes in the phase and amplitude of the transmitted signal due to the effect of multipath propagation, i.e., the presence of multiple signal components that propagate over different paths. 
It provides the possibility to set the model parameters depending on the scenario of interest, thus enabling the modeling of multiple propagation environments.

Eq.~\eqref{eq:channel} represents the overall channel impulse response $H_{u,s}(t,\tau)$. As mentioned in Section~\ref{sec:channel}, it is obtained by the superposition of $M \times N$ rays, grouped in $N$ clusters. Rays belonging to the same cluster experience the same power $P_n$ and propagation delay $\tau_n$, present similar angles of arrival ($\theta_{n,m}^A, \phi_{n,m}^A$) and departure ($\theta_{n,m}^D, \phi_{n,m}^D$), and have uniformly distributed initial phases $\Phi_{n,m}$.
Each ray accounts for the antenna field patterns $\mathbf{F} (\theta_{n,m}, \phi_{n,m})$ and for the power distribution among the vertical and horizontal polarizations through the term $K_{n,m}$.
The terms $exp (j\bar{\mathbf{k}}^T \bar{\mathbf{d}})$ represent the array responses of the transmitting and receiving antennas, where $\bar{\mathbf{k}}$ is the wave vector and $\bar{\mathbf{d}}$ is the element location vector.
In case of user mobility, each ray is subject to a phase shift $\nu_{n,m}$ due to the Doppler effect.
In the \gls{los} case, a Ricean factor is added to the direct path.

\begin{equation}\label{eq:channel}
  \begin {split}
  H_{u,s} (t,\tau)
   =&\sum_{n=1}^N\sqrt{\frac{P_n}{M}}\sum_{m=1}^M\bar{\mathbf{F}}_{rx} (\theta_{n,m}^A, \phi_{n,m}^A) \\
  &\times\begin{bmatrix} e^{j\Phi_{n,m}^{\theta, \theta}} & \sqrt{K^{-1}_{n,m}}e^{j\Phi_{n,m}^{\theta, \phi}} \\ \sqrt{K^{-1}_{n,m}}e^{j\Phi_{n,m}^{\phi, \theta}} & e^{j\Phi_{n,m}^{\phi, \phi}} \end{bmatrix}\\
  &\times\bar{\mathbf{F}}_{tx} (\theta_{n,m}^D, \phi_{n,m}^D)\\
  &\times e^{j\bar{\mathbf{k}}_{rx,n,m}^T \bar{\mathbf{d}}_{rx,u}}
  e^{j\bar{\mathbf{k}}_{tx,n,m}^T \bar{\mathbf{d}}_{tx,s}} \\
  &\times e^{j2\pi \nu_{n,m}t}\delta (\tau - \tau_n)
  \end{split}
\end{equation}


Our implementation follows the same approach described in \cite{zhang2017ns3}, but introduces some changes to improve the modularity of the code and includes the latest updates with respect to \cite{38900}.
As in \cite{zhang2017ns3}, to reduce the model complexity, we assumed that all rays within a cluster are subject to the same Doppler shift ($\nu_n$), corresponding to that of the central ray.
Thus, the channel impulse response can be expressed as:
\begin{equation}\label{eq:channel2}
  H_{u,s} (t,\tau) = \sum_{n=1}^N H_{u,s,n} e^{j2\pi \nu_{n}t} \delta (\tau - \tau_n),
\end{equation}
where $H_{u,s,n}$ represents all the terms of the impulse response except for the Doppler contribution.

We developed the class \texttt{ThreeGppChannelModel}, which computes the coefficients $H_{u,s,n}$ as described in Section~7.5 of \cite{TR38901} and handles their periodic update.
It is associated with an instance of \texttt{ChannelConditionModel}, used to determine the \gls{los}/\gls{nlos} channel state.
The main method is \texttt{GetChannel}, which takes as input the mobility models of the transmitter and receiver nodes and the associated antenna objects, and returns an instance of \texttt{ThreeGppChannelMatrix}.
As represented in Table~\ref{tab:matrix}, the structure \texttt{ThreeGppChannelMatrix} contains entries to store the channel coefficients $H_{u,s,n}$, the propagation delays $\tau_n$, the angles of arrival and departure, and a time stamp indicating the generation time.
\begin{table}
\centering
\caption{Main Entries of \texttt{ThreeGppChannelMatrix}}\label{tab:matrix}
\vspace{-0.2cm}
\small
\begin{tabular}{|ll|}
  \hline
  \multicolumn{2}{|c|}{\textbf{ThreeGppChannelMatrix}}\\
  \hline
  \texttt{m\_channel} & the channel coefficients $H_{u,s,n}$\\
  \texttt{m\_delay} & the clusters delays $\tau_n$\\
  \texttt{m\_angle} & the clusters arrival and departure angles\\
  \texttt{m\_generatedTime} & a time stamp indicating the generation time\\
  \texttt{m\_nodeIds} & IDs of the transmitter and receiver nodes\\
\hline
\end{tabular}
\vspace{-0.5cm}
\end{table}
The first time a channel is generated, the corresponding \texttt{ThreeGppChannelMatrix} is cached in a map together with identifiers for the transmitting and receiving nodes. When the same channel is requested again, the method \texttt{GetChannel} retrieves the \texttt{ThreeGppChannelMatrix} from the map and checks whether the channel coefficients have to be updated or not, depending on the expired time and the occurrence of \gls{los}-\gls{nlos} transitions. If so, it recomputes the coefficients, otherwise it returns the old realization.
Moreover, the class \texttt{ThreeGppChannelModel} provides attributes to enable an easy configuration of the model parameters, such as carrier frequency, channel scenario and update period.
In particular, the choice of the update period should consider (i) the channel coherence time, i.e., the time duration over which the channel response does not vary, which depends on several factors, such as frequency, user mobility and propagation environment, and (ii) the time granularity of the simulation, which should be fine enough to capture the channel dynamics.

\subsubsection{Blockage Model}
3GPP TR 38.901 also provides an optional feature that can be used to model the blockage effect due to the presence of obstacles, such as trees, cars or humans, at the level of a single cluster. This differs from a complete blockage, 
which would result in an \gls{los} to \gls{nlos} transition.
Therefore, when this feature is enabled, an additional attenuation is added to certain clusters, depending on their angle of arrival.
There are two possible methods for the computation of the additional attenuation, i.e., stochastic (Model A) and geometric (Model B).
In this work, we used the implementation provided by Zhang et al. in \cite{zhang2017ns3}, which uses the stochastic method.
In particular, we extended the class \texttt{ThreeGppChannelModel} by including the method \texttt{CalcAttenuationOfBlockage}, which computes the additional  attenuation.
Also, we defined attributes to enable/disable the blockage feature and to configure the model parameters.

\vspace{-0.2cm}
\subsection{Antenna Array Model}
In 3GPP TR 38.901, BS and UT antennas are modeled as uniform rectangular antenna arrays with multiple panels, each containing $N_{a,c} \times N_{a,r}$ antenna elements with fixed spacing, where $N_{a,c}$ is the number of elements in a column and $N_{a,r}$ is the number of elements in a row of the antenna array. Orientation of the arrays can be configured by adjusting bearing, tilt and slant angles.

The technical report \cite{TR38901} describes how to compute the azimuth and zenith components of the field pattern $\mathbf{F}$ to be used for the generation of the channel matrix, i.e., $F_{\theta}$ and $F_{\phi}$, respectively, taking into account orientation and polarization of the panels.
We developed a new class, called \texttt{ThreeGppAntennaArrayModel}, which implements this model under the following assumptions: (i) the array is composed of a single panel; (ii) the slant angle is fixed and equal to $0$ degrees; and (iii) the antenna elements are vertically polarized.

The main method is \texttt{GetElementFieldPattern}, which accepts as argument the azimuth and zenith angles of arrival and returns \texttt{std::pair} containing the element field components computed following a four-step procedure.
The first step is to express the azimuth and zenith angles of arrival using the antenna local coordinate system, i.e., with the origin corresponding to the bottom-left corner of the array and x- and y-axes parallel to the horizontal and vertical sides, using the procedure described in Section~7.1 of \cite{TR38901}.
The second step is to compute the radiation power pattern of each element, which describes how the irradiated power varies in space.
The technical report specifies the radiation power pattern as a function of the zenith and azimuth angles of arrival $(\theta', \phi')$, expressed in the local coordinate system: 
\begin{equation}
  \begin{split}
A'_{dB} (\theta', \phi') = &G_E - \min \Bigg{\{} \min \bigg{[} 12 \bigg{(} \frac{\theta' - 90^\circ}{\theta_{3dB}}\bigg{)}^{2} , SLA_V \bigg{]} \\
& - \min \bigg{[} 12 \bigg{(} \frac{\phi'}{\phi_{3dB}}\bigg{)}^{2} , A_{\max} \bigg{]}, A_{\max} \Bigg{\}},
\end{split}
\end{equation}
where $\theta_{3dB}$ and $\phi_{3dB}$ represent the vertical and horizontal beamwidth and are equal to $65^\circ$, $SLA_V$ is the side-lobe level limit and is set to $30$~dB, $A_{max}$ represents the front-back ratio and is fixed to $30$~dB, while $G_E$ represents the directional gain of an antenna element and is a model parameter.
The third step is to derive the azimuth and zenith components of the element field pattern. Since only vertical polarization is considered, we have $F'_{\phi'}(\theta', \phi') = 0$ and $F'_{\theta'}(\theta', \phi') = \sqrt{A'(\theta', \phi')}$.
Finally, the fourth step is to convert the field pattern components back to the global coordinate system to obtain $F_{\theta}(\theta, \phi)$ and $F_{\phi}(\theta, \phi)$.

The \texttt{ThreeGppAntennaArrayModel} class features also the method \texttt{GetElementLocation}, which accepts as argument the index of the antenna element and returns the corresponding location vector $\bar{\mathbf{d}}$, and provides the methods \texttt{Set/GetBeamformingVector} to store and retrieve the beamforming vector $\mathbf{w}$.


\vspace{-0.2cm}
\subsection{Computation of the \gls{psd}}
The \gls{psd} of the received signal is computed as:
\begin{equation}\label{eq:psd}
  S_{rx} (t, f) = S_{tx} (t, f) \mathbf{w}_{rx}^T \mathcal{H} (t, f) \mathbf{w}_{tx} ,
\end{equation}
where $S_{tx} (t, f)$ is the \gls{psd} of the transmitted signal, $\mathbf{w}_{rx}$ and $\mathbf{w}_{tx}$ are the transmitting and receiving beamforming vectors, and $\mathcal{H} (t, f)$ is the channel matrix in the frequency domain.
Applying the Fourier transform to channel coefficients expressed as in Eq. \eqref{eq:channel2}, $S_{rx} (t, f)$ can be rewritten as:
\begin{equation}\label{eq:psd2}
  \begin{split}
  S_{rx} (t, f) &= S_{tx} (t, f) \sum_{n=1}^N \sum_{s=1}^S \sum_{u=1}^U w_{rx, u} H_{u,s,n} w_{tx, s} e^{j2\pi \nu_{n}t} e^{j2\pi\tau_n f}\\
  &= S_{tx} (t, f) \sum_{n=1}^N L_{n} e^{j2\pi \nu_{n}t} e^{j2\pi\tau_n f} ,
  \end{split}
\end{equation}
where $L_{n}$ represents the long-term component of cluster $n$, as defined in \cite{zhang2017ns3}.

In our implementation, the computation of $S_{rx}(t, f)$ is handled by the class \texttt{ThreeGppSpectrumPropagationLossModel}, which extends the \texttt{SpectrumPropagationLossModel} interface.
This class interacts with \texttt{ThreeGppChannelModel} to retrieve the channel coefficients and holds a map containing the \texttt{ThreeGppAntennaArray} objects of all the devices.
The main method is \texttt{DoCalcRxPowerSpectralDensity}, which takes as input the mobility models of transmitter and receiver nodes, and returns the \gls{psd} of the received signal, computed using Eq.~\eqref{eq:psd2}.
In particular, it relies on the private methods \texttt{GetLongTerm}, to calculate the long term components, and \texttt{CalcBeamformingGain}, to account for the Doppler and the propagation delay.
To reduce the computational load, all the long term components associated with a certain channel are cached and recomputed only when the channel realization is updated.
Also, \texttt{ThreeGppSpectrumPropagationLossModel} provides the method \texttt{SetChannelModelAttribute}, which can be used to configure the model parameters, such as carrier frequency and channel scenario.
\vspace{-0.2cm}
\section{Examples and Use Cases}
\label{sec:scenarios}
In this section, we describe, by means of an example, how the different classes presented above have to be configured to build the entire 3GPP channel modeling framework.
Also, we outline some possible use cases.

\begin{figure*}
\setlength\belowcaptionskip{-0.2cm}
\begin{subfigure}{0.245\textwidth}
	\centering
  \setlength\abovecaptionskip{0.1cm}
  \setlength\belowcaptionskip{-0.33cm}
	\setlength\fwidth{1.1\columnwidth}
	\setlength\fheight{0.8\columnwidth}
\begin{tikzpicture}

\definecolor{color0}{rgb}{0.12156862745098,0.466666666666667,0.705882352941177}
\definecolor{color1}{rgb}{1,0.498039215686275,0.0549019607843137}
\pgfplotsset{every tick label/.append style={font=\scriptsize}}

\begin{axis}[
width=0.951\fwidth,
height=\fheight,
at={(0\fwidth,0\fheight)},
legend cell align={left},
legend style={at={(0.03,0.97)}, anchor=north west, draw=white!80.0!black},
tick align=inside,
tick pos=left,
x grid style={white!69.01960784313725!black},
xlabel={Distance [m]},
xmin=0, xmax=4950,
xtick style={color=black},
y grid style={white!69.01960784313725!black},
ylabel={Average Loss [dB]},
ymin=60.30259965, ymax=170.07497335,
ytick style={color=black},
ylabel style={font=\footnotesize\color{white!15!black}},
xlabel style={font=\footnotesize\color{white!15!black}},
ymajorgrids,
xmajorgrids,
]
\addplot [semithick, color0, mark repeat=15, mark=*, mark size={1.5}]
table {%
50 75.116008
99.5 79.969878
149 83.114159
198.5 85.153225
248 88.257605
297.5 89.168451
347 90.901291
396.5 91.456324
446 93.017537
495.5 94.240582
545 95.136613
594.5 96.441561
644 96.725363
693.5 97.411082
743 98.371452
792.5 98.893525
842 100.064771
891.5 100.476736
941 100.108924
990.5 100.458691
1040 101.68603
1089.5 101.861872
1139 101.678143
1188.5 102.878391
1238 103.52586
1287.5 104.395167
1337 104.699755
1386.5 105.50308
1436 105.018291
1485.5 104.94436
1535 105.189206
1584.5 106.951601
1634 107.081619
1683.5 106.467112
1733 107.257175
1782.5 107.441989
1832 107.797477
1881.5 107.833143
1931 108.013431
1980.5 108.352887
2030 108.37222
2079.5 108.448091
2129 109.08836
2178.5 108.79829
2228 110.30947
2277.5 110.32233
2327 109.62063
2376.5 111.05903
2426 111.196009
2475.5 111.493974
2525 111.885283
2574.5 112.019521
2624 112.547018
2673.5 111.819769
2723 113.76531
2772.5 114.279576
2822 114.317724
2871.5 113.502072
2921 112.074644
2970.5 113.67078
3020 115.04177
3069.5 115.77287
3119 115.77506
3168.5 114.319015
3218 115.126272
3267.5 115.8023
3317 116.42143
3366.5 116.22517
3416 115.92093
3465.5 116.29825
3515 116.81763
3564.5 117.53531
3614 118.20422
3663.5 118.97008
3713 118.67542
3762.5 118.49084
3812 118.58857
3861.5 119.39043
3911 118.99986
3960.5 119.25738
4010 119.62142
4059.5 120.37304
4109 120.2986
4158.5 120.32796
4208 120.78325
4257.5 121.4173
4307 120.81847
4356.5 120.7536
4406 121.61052
4455.5 122.12732
4505 121.99539
4554.5 121.76506
4604 122.64368
4653.5 121.83169
4703 121.79738
4752.5 122.39009
4802 122.47778
4851.5 124.09423
4901 122.98227
4950.5 123.4212
};
\addplot [semithick, color1, mark repeat=15, mark=triangle*]
table {%
50 77.756507
99.5 88.496122
149 94.298105
198.5 99.584573
248 103.544974
297.5 106.324968
347 107.788247
396.5 110.24939
446 112.028426
495.5 113.33521
545 114.591973
594.5 116.953199
644 117.427119
693.5 119.69572
743 121.44954
792.5 122.53775
842 123.66143
891.5 125.45699
941 126.39047
990.5 127.2884
1040 128.395382
1089.5 129.23082
1139 129.69702
1188.5 130.30144
1238 130.40211
1287.5 130.76698
1337 130.746599
1386.5 131.011539
1436 131.53977
1485.5 132.10108
1535 132.90652
1584.5 134.07656
1634 135.1006
1683.5 134.27587
1733 134.54714
1782.5 135.14903
1832 136.00463
1881.5 136.21606
1931 137.26607
1980.5 138.69043
2030 139.38855
2079.5 139.62661
2129 140.14414
2178.5 139.93236
2228 139.65929
2277.5 140.06242
2327 140.70591
2376.5 141.9087
2426 141.70022
2475.5 141.3655
2525 141.8961
2574.5 141.5299
2624 141.30461
2673.5 141.84704
2723 142.47878
2772.5 142.67846
2822 143.05009
2871.5 143.69303
2921 145.02717
2970.5 145.98883
3020 144.87624
3069.5 146.25677
3119 144.65276
3168.5 144.66214
3218 145.15861
3267.5 145.49404
3317 146.44225
3366.5 146.14322
3416 146.85674
3465.5 146.48841
3515 146.50673
3564.5 147.03729
3614 147.75074
3663.5 147.9541
3713 147.37326
3762.5 147.37072
3812 147.52156
3861.5 149.116
3911 149.69646
3960.5 149.3385
4010 149.32216
4059.5 147.57474
4109 149.48134
4158.5 148.99151
4208 150.02265
4257.5 151.2043
4307 151.15358
4356.5 151.24338
4406 151.86272
4455.5 150.94617
4505 150.7221
4554.5 150.00375
4604 149.56418
4653.5 150.94778
4703 151.54643
4752.5 151.03888
4802 150.68959
4851.5 150.65587
4901 151.7509
4950.5 151.84996
};

\addplot [semithick, dashed, black]
table {%
50 74.8041
99.5 79.8208
149 83.2218
198.5 85.7469
248 87.7511
297.5 89.4141
347 90.837
396.5 92.0823
446 93.191
495.5 94.1913
545 95.1037
594.5 95.9433
644 96.7216
693.5 97.4476
743 98.1286
792.5 98.7702
842 99.3773
891.5 99.9537
941 100.503
990.5 101.027
1040 101.53
1089.5 102.012
1139 102.476
1188.5 102.923
1238 103.355
1287.5 103.773
1337 104.177
1386.5 104.569
1436 104.95
1485.5 105.321
1535 105.681
1584.5 106.033
1634 106.375
1683.5 106.71
1733 107.037
1782.5 107.356
1832 107.669
1881.5 107.975
1931 108.275
1980.5 108.569
2030 108.858
2079.5 109.141
2129 109.42
2178.5 109.693
2228 109.962
2277.5 110.227
2327 110.487
2376.5 110.744
2426 110.996
2475.5 111.245
2525 111.555
2574.5 111.892
2624 112.223
2673.5 112.548
2723 112.867
2772.5 113.179
2822 113.487
2871.5 113.789
2921 114.086
2970.5 114.378
3020 114.665
3069.5 114.947
3119 115.225
3168.5 115.498
3218 115.768
3267.5 116.033
3317 116.294
3366.5 116.551
3416 116.805
3465.5 117.055
3515 117.301
3564.5 117.544
3614 117.784
3663.5 118.02
3713 118.253
3762.5 118.483
3812 118.71
3861.5 118.934
3911 119.155
3960.5 119.374
4010 119.59
4059.5 119.803
4109 120.013
4158.5 120.221
4208 120.427
4257.5 120.63
4307 120.831
4356.5 121.029
4406 121.226
4455.5 121.42
4505 121.612
4554.5 121.801
4604 121.989
4653.5 122.175
4703 122.359
4752.5 122.541
4802 122.721
4851.5 122.899
4901 123.075
4950.5 123.25
};

\addplot [semithick, dashed, black]
table {%
50 78.6867
99.5 88.0329
149 94.3233
198.5 98.959
248 102.611
297.5 105.619
347 108.173
396.5 110.393
446 112.354
495.5 114.111
545 115.702
594.5 117.156
644 118.494
693.5 119.733
743 120.887
792.5 121.968
842 122.982
891.5 123.939
941 124.845
990.5 125.704
1040 126.521
1089.5 127.301
1139 128.046
1188.5 128.759
1238 129.443
1287.5 130.1
1337 130.733
1386.5 131.342
1436 131.931
1485.5 132.499
1535 133.049
1584.5 133.581
1634 134.097
1683.5 134.597
1733 135.083
1782.5 135.556
1832 136.015
1881.5 136.462
1931 136.898
1980.5 137.322
2030 137.737
2079.5 138.141
2129 138.535
2178.5 138.921
2228 139.298
2277.5 139.666
2327 140.027
2376.5 140.38
2426 140.726
2475.5 141.065
2525 141.397
2574.5 141.723
2624 142.042
2673.5 142.356
2723 142.663
2772.5 142.966
2822 143.262
2871.5 143.554
2921 143.841
2970.5 144.123
3020 144.4
3069.5 144.673
3119 144.941
3168.5 145.205
3218 145.465
3267.5 145.721
3317 145.974
3366.5 146.222
3416 146.467
3465.5 146.708
3515 146.946
3564.5 147.181
3614 147.412
3663.5 147.641
3713 147.866
3762.5 148.088
3812 148.307
3861.5 148.524
3911 148.737
3960.5 148.948
4010 149.157
4059.5 149.363
4109 149.566
4158.5 149.767
4208 149.965
4257.5 150.162
4307 150.355
4356.5 150.547
4406 150.737
4455.5 150.924
4505 151.11
4554.5 151.293
4604 151.474
4653.5 151.654
4703 151.831
4752.5 152.007
4802 152.181
4851.5 152.353
4901 152.523
4950.5 152.692
};
\end{axis}

\end{tikzpicture}
  \caption{RMa}
\end{subfigure}%
\hfill%
\begin{subfigure}{0.245\textwidth}
	\centering
  \setlength\abovecaptionskip{0.1cm}
  \setlength\belowcaptionskip{-0.33cm}
	\setlength\fwidth{1.1\columnwidth}
	\setlength\fheight{0.8\columnwidth}
\begin{tikzpicture}

\definecolor{color0}{rgb}{0.12156862745098,0.466666666666667,0.705882352941177}
\definecolor{color1}{rgb}{1,0.498039215686275,0.0549019607843137}
\pgfplotsset{every tick label/.append style={font=\scriptsize}}

\begin{axis}[
width=0.951\fwidth,
height=\fheight,
at={(0\fwidth,0\fheight)},
legend cell align={left},
legend style={at={(0.99,0.01)}, font=\footnotesize, anchor=south east, draw=white!80.0!black},
tick align=inside,
tick pos=left,
x grid style={white!69.01960784313725!black},
xlabel={Distance [m]},
xmin=0, xmax=4950,
xtick style={color=black},
y grid style={white!69.01960784313725!black},
ymin=60.30259965, ymax=170.07497335,
ytick style={color=black},
ylabel style={font=\footnotesize\color{white!15!black}},
xlabel style={font=\footnotesize\color{white!15!black}},
ymajorgrids,
xmajorgrids,
legend columns=1
]
\addplot [semithick, color0, mark repeat=15, mark=*, mark size={1.5}]
table {%
10 65.292253
59.9 73.94262
109.8 79.09662
159.7 83.68966
209.6 85.642744
259.5 87.829582
309.4 89.206507
359.3 90.192834
409.2 92.650011
459.1 94.64272
509 96.6193
558.9 98.5586
608.8 99.354089
658.7 99.842027
708.6 101.531497
758.5 102.353404
808.4 103.751478
858.3 104.46718
908.2 105.409074
958.1 106.444083
1008 107.520073
1057.9 108.81894
1107.8 109.65674
1157.7 109.736896
1207.6 111.00122
1257.5 111.654253
1307.4 112.6294
1357.3 113.39385
1407.2 113.45062
1457.1 114.52526
1507 114.9561
1556.9 114.80628
1606.8 115.65466
1656.7 116.56588
1706.6 116.88901
1756.5 118.05061
1806.4 118.72588
1856.3 118.71426
1906.2 118.83745
1956.1 119.75209
2006 119.65033
2055.9 119.86412
2105.8 120.21095
2155.7 120.67152
2205.6 121.15621
2255.5 122.47754
2305.4 123.29807
2355.3 123.63761
2405.2 123.68806
2455.1 123.33043
2505 123.49736
2554.9 123.51309
2604.8 124.61438
2654.7 124.00194
2704.6 124.31743
2754.5 125.68217
2804.4 124.90972
2854.3 125.26025
2904.2 126.14858
2954.1 125.87055
3004 126.85615
3053.9 127.14884
3103.8 127.25503
3153.7 127.68874
3203.6 127.69632
3253.5 127.65472
3303.4 127.19816
3353.3 128.39779
3403.2 128.96801
3453.1 129.78149
3503 129.07733
3552.9 129.32289
3602.8 130.44535
3652.7 129.64069
3702.6 130.49312
3752.5 130.95973
3802.4 130.82684
3852.3 130.91126
3902.2 130.98064
3952.1 130.86699
4002 131.51301
4051.9 132.264
4101.8 131.97944
4151.7 132.91173
4201.6 133.28521
4251.5 133.52153
4301.4 133.45503
4351.3 132.74527
4401.2 133.13466
4451.1 132.82139
4501 133.23396
4550.9 133.25236
4600.8 133.58947
4650.7 133.96128
4700.6 134.27925
4750.5 134.79762
4800.4 135.37176
4850.3 134.92637
4900.2 134.78262
4950.1 135.71582
};
\addlegendentry{LOS}
\addplot [semithick, color1, mark repeat=15, mark=triangle*]
table {%
10 75.122151
59.9 89.566905
109.8 100.403682
159.7 105.517537
209.6 110.732006
259.5 114.424854
309.4 117.48071
359.3 119.52288
409.2 121.91576
459.1 123.9755
509 125.48733
558.9 128.06621
608.8 129.44162
658.7 129.56826
708.6 131.87622
758.5 132.11795
808.4 133.23633
858.3 134.65961
908.2 136.25198
958.1 135.86825
1008 136.45705
1057.9 138.75097
1107.8 139.68383
1157.7 139.88909
1207.6 140.73557
1257.5 140.61187
1307.4 141.10434
1357.3 143.1221
1407.2 142.93574
1457.1 143.13864
1507 142.91093
1556.9 143.9073
1606.8 145.48499
1656.7 146.02363
1706.6 146.73398
1756.5 147.51787
1806.4 147.47178
1856.3 148.47991
1906.2 150.18961
1956.1 149.49755
2006 148.7405
2055.9 149.99443
2105.8 150.78423
2155.7 150.47687
2205.6 151.02756
2255.5 150.09024
2305.4 151.34926
2355.3 151.98571
2405.2 152.96563
2455.1 153.6242
2505 153.41635
2554.9 153.37819
2604.8 153.11266
2654.7 153.45167
2704.6 153.98445
2754.5 153.64517
2804.4 155.08688
2854.3 155.05194
2904.2 155.02937
2954.1 155.6209
3004 156.29628
3053.9 156.74828
3103.8 156.05528
3153.7 155.84106
3203.6 157.59291
3253.5 157.95095
3303.4 157.74969
3353.3 157.35653
3403.2 157.90874
3453.1 159.71
3503 158.71426
3552.9 158.41821
3602.8 159.09357
3652.7 158.49739
3702.6 159.55137
3752.5 159.5224
3802.4 159.62508
3852.3 159.2924
3902.2 159.82036
3952.1 160.5344
4002 160.38386
4051.9 161.72724
4101.8 161.70218
4151.7 161.51741
4201.6 162.71888
4251.5 161.61015
4301.4 161.28632
4351.3 161.25003
4401.2 160.97743
4451.1 162.67095
4501 162.99844
4550.9 162.25594
4600.8 163.71806
4650.7 163.81547
4700.6 164.0241
4750.5 164.54985
4800.4 164.70485
4850.3 164.70575
4900.2 165.08532
4950.1 164.64525
};
\addlegendentry{NLOS}

\addplot [semithick, dashed, black]
table {%
10 65.4713
59.9 74.3291
109.8 79.6526
159.7 83.1214
209.6 85.677
259.5 87.6969
309.4 89.3659
359.3 90.7875
409.2 92.0482
459.1 94.0413
509 95.8295
558.9 97.451
608.8 98.9343
658.7 100.301
708.6 101.568
758.5 102.749
808.4 103.855
858.3 104.894
908.2 105.875
958.1 106.804
1008 107.686
1057.9 108.524
1107.8 109.325
1157.7 110.09
1207.6 110.823
1257.5 111.526
1307.4 112.202
1357.3 112.852
1407.2 113.479
1457.1 114.084
1507 114.669
1556.9 115.235
1606.8 115.783
1656.7 116.314
1706.6 116.829
1756.5 117.33
1806.4 117.816
1856.3 118.29
1906.2 118.751
1956.1 119.199
2006 119.637
2055.9 120.064
2105.8 120.48
2155.7 120.887
2205.6 121.285
2255.5 121.673
2305.4 122.053
2355.3 122.425
2405.2 122.789
2455.1 123.146
2505 123.496
2554.9 123.838
2604.8 124.174
2654.7 124.504
2704.6 124.827
2754.5 125.145
2804.4 125.457
2854.3 125.763
2904.2 126.064
2954.1 126.36
3004 126.651
3053.9 126.937
3103.8 127.219
3153.7 127.496
3203.6 127.768
3253.5 128.037
3303.4 128.301
3353.3 128.562
3403.2 128.818
3453.1 129.071
3503 129.32
3552.9 129.566
3602.8 129.808
3652.7 130.047
3702.6 130.283
3752.5 130.516
3802.4 130.745
3852.3 130.972
3902.2 131.195
3952.1 131.416
4002 131.634
4051.9 131.849
4101.8 132.062
4151.7 132.272
4201.6 132.479
4251.5 132.684
4301.4 132.887
4351.3 133.087
4401.2 133.286
4451.1 133.481
4501 133.675
4550.9 133.867
4600.8 134.056
4650.7 134.243
4700.6 134.429
4750.5 134.612
4800.4 134.794
4850.3 134.973
4900.2 135.151
4950.1 135.327
};

\addplot [semithick, dashed, black]
table {%
10 74.9596
59.9 90.6943
109.8 100.151
159.7 106.313
209.6 110.852
259.5 114.44
309.4 117.405
359.3 119.93
409.2 122.129
459.1 124.077
509 125.824
558.9 127.408
608.8 128.857
658.7 130.192
708.6 131.43
758.5 132.584
808.4 133.664
858.3 134.68
908.2 135.639
958.1 136.546
1008 137.407
1057.9 138.227
1107.8 139.009
1157.7 139.756
1207.6 140.472
1257.5 141.159
1307.4 141.819
1357.3 142.455
1407.2 143.067
1457.1 143.659
1507 144.23
1556.9 144.783
1606.8 145.318
1656.7 145.837
1706.6 146.34
1756.5 146.83
1806.4 147.305
1856.3 147.767
1906.2 148.217
1956.1 148.656
2006 149.083
2055.9 149.5
2105.8 149.907
2155.7 150.305
2205.6 150.693
2255.5 151.073
2305.4 151.444
2355.3 151.808
2405.2 152.163
2455.1 152.512
2505 152.853
2554.9 153.188
2604.8 153.516
2654.7 153.838
2704.6 154.154
2754.5 154.465
2804.4 154.769
2854.3 155.069
2904.2 155.363
2954.1 155.652
3004 155.936
3053.9 156.216
3103.8 156.491
3153.7 156.762
3203.6 157.028
3253.5 157.29
3303.4 157.549
3353.3 157.803
3403.2 158.054
3453.1 158.301
3503 158.544
3552.9 158.784
3602.8 159.021
3652.7 159.254
3702.6 159.485
3752.5 159.712
3802.4 159.936
3852.3 160.157
3902.2 160.376
3952.1 160.592
4002 160.804
4051.9 161.015
4101.8 161.222
4151.7 161.428
4201.6 161.63
4251.5 161.831
4301.4 162.029
4351.3 162.225
4401.2 162.418
4451.1 162.61
4501 162.799
4550.9 162.986
4600.8 163.171
4650.7 163.354
4700.6 163.535
4750.5 163.714
4800.4 163.892
4850.3 164.067
4900.2 164.241
4950.1 164.413
};

\end{axis}

\end{tikzpicture}
  \caption{UMa}
\end{subfigure}%
\hfill%
\begin{subfigure}{0.245\textwidth}
	\centering
  \setlength\abovecaptionskip{0.1cm}
  \setlength\belowcaptionskip{-0.33cm}
	\setlength\fwidth{1.1\columnwidth}
	\setlength\fheight{0.8\columnwidth}
\begin{tikzpicture}

\definecolor{color0}{rgb}{0.12156862745098,0.466666666666667,0.705882352941177}
\definecolor{color1}{rgb}{1,0.498039215686275,0.0549019607843137}
\pgfplotsset{every tick label/.append style={font=\scriptsize}}

\begin{axis}[
width=0.951\fwidth,
height=\fheight,
at={(0\fwidth,0\fheight)},
legend cell align={left},
legend style={at={(-0.5,1.02)}, anchor=north west, draw=white!80.0!black},
tick align=inside,
tick pos=left,
x grid style={white!69.01960784313725!black},
xlabel={Distance [m]},
xmin=0, xmax=4950,
xtick style={color=black},
y grid style={white!69.01960784313725!black},
ymin=60.30259965, ymax=170.07497335,
ytick style={color=black},
ylabel style={font=\footnotesize\color{white!15!black}},
xlabel style={font=\footnotesize\color{white!15!black}},
legend columns=2,
xmajorgrids,
ymajorgrids,
]
\addplot [semithick, color0, mark repeat=15, mark=*, mark size={1.5}]
table {%
10 61.624414
59.9 76.428489
109.8 82.518947
159.7 86.388584
209.6 89.66427
259.5 93.837996
309.4 97.172887
359.3 100.184501
409.2 101.942181
459.1 104.046054
509 105.737827
558.9 107.331006
608.8 108.948007
658.7 110.37561
708.6 111.46669
758.5 113.32632
808.4 113.94413
858.3 114.76176
908.2 115.2257
958.1 116.59611
1008 117.48629
1057.9 118.38485
1107.8 119.43585
1157.7 119.5912
1207.6 120.47125
1257.5 121.26819
1307.4 122.2711
1357.3 123.14021
1407.2 123.83794
1457.1 123.88312
1507 124.07344
1556.9 125.27125
1606.8 125.53814
1656.7 126.0387
1706.6 126.91064
1756.5 126.80775
1806.4 127.56428
1856.3 128.0035
1906.2 128.65299
1956.1 129.19071
2006 129.46628
2055.9 129.97461
2105.8 129.95176
2155.7 130.84869
2205.6 131.71387
2255.5 131.06821
2305.4 131.71095
2355.3 133.03673
2405.2 132.95844
2455.1 132.781
2505 132.69091
2554.9 133.16274
2604.8 133.43708
2654.7 134.357
2704.6 134.50737
2754.5 134.94103
2804.4 135.38883
2854.3 135.31092
2904.2 136.05043
2954.1 135.80323
3004 136.87697
3053.9 136.37628
3103.8 137.19945
3153.7 136.856
3203.6 137.37901
3253.5 138.52689
3303.4 137.56301
3353.3 138.94043
3403.2 139.04716
3453.1 139.11875
3503 138.78635
3552.9 139.7493
3602.8 139.89934
3652.7 139.26801
3702.6 140.33837
3752.5 140.41869
3802.4 141.27615
3852.3 140.69815
3902.2 141.27278
3952.1 140.7597
4002 141.37569
4051.9 141.60159
4101.8 141.79711
4151.7 142.77546
4201.6 142.44691
4251.5 142.23019
4301.4 142.28575
4351.3 143.31218
4401.2 143.34783
4451.1 143.30666
4501 142.8122
4550.9 142.76604
4600.8 144.09574
4650.7 144.22962
4700.6 143.5109
4750.5 144.93163
4800.4 144.93722
4850.3 145.00536
4900.2 145.32377
4950.1 145.85178
};

\addplot [semithick, color1, mark repeat=15, mark=triangle*]
table {%
10 67.247256
59.9 92.571839
109.8 101.855518
159.7 106.813213
209.6 111.444037
259.5 115.390664
309.4 117.106769
359.3 118.37514
409.2 121.05691
459.1 122.60029
509 124.488971
558.9 125.78106
608.8 125.96611
658.7 127.91425
708.6 129.23241
758.5 130.51413
808.4 131.52365
858.3 132.7708
908.2 134.80761
958.1 134.96567
1008 136.31237
1057.9 134.56476
1107.8 136.33104
1157.7 136.44237
1207.6 137.33747
1257.5 138.60588
1307.4 138.14593
1357.3 139.69516
1407.2 140.48819
1457.1 141.72378
1507 141.90127
1556.9 142.28598
1606.8 142.00642
1656.7 143.02535
1706.6 143.26241
1756.5 143.38982
1806.4 144.02606
1856.3 144.77193
1906.2 144.68343
1956.1 145.64735
2006 147.08347
2055.9 146.92389
2105.8 146.75829
2155.7 146.61776
2205.6 147.40026
2255.5 148.1413
2305.4 147.03087
2355.3 149.12995
2405.2 148.40674
2455.1 149.08667
2505 149.25493
2554.9 147.97916
2604.8 148.84817
2654.7 150.13447
2704.6 149.61714
2754.5 151.41962
2804.4 150.87454
2854.3 151.72542
2904.2 151.68693
2954.1 151.97986
3004 151.68033
3053.9 151.37864
3103.8 153.76029
3153.7 152.78463
3203.6 154.26404
3253.5 153.96908
3303.4 152.34825
3353.3 152.62581
3403.2 152.97226
3453.1 153.68282
3503 152.96014
3552.9 154.29562
3602.8 153.8898
3652.7 155.05955
3702.6 154.96667
3752.5 155.96242
3802.4 155.76502
3852.3 157.27645
3902.2 156.53471
3952.1 155.72083
4002 157.58776
4051.9 156.43765
4101.8 157.84172
4151.7 155.97505
4201.6 157.06332
4251.5 157.85454
4301.4 158.4369
4351.3 157.23141
4401.2 158.00222
4451.1 158.44534
4501 158.51045
4550.9 157.00734
4600.8 158.87702
4650.7 160.05374
4700.6 158.87649
4750.5 159.50833
4800.4 157.66836
4850.3 160.12311
4900.2 159.98984
4950.1 159.07663
};

\addplot [semithick, dashed, black]
table {%
10 62.3819
59.9 76.3619
109.8 81.8264
159.7 85.5818
209.6 90.2952
259.5 94.0002
309.4 97.0528
359.3 99.6487
409.2 101.907
459.1 103.905
509 105.697
558.9 107.321
608.8 108.806
658.7 110.175
708.6 111.443
758.5 112.625
808.4 113.732
858.3 114.772
908.2 115.754
958.1 116.683
1008 117.565
1057.9 118.404
1107.8 119.205
1157.7 119.97
1207.6 120.703
1257.5 121.406
1307.4 122.082
1357.3 122.733
1407.2 123.36
1457.1 123.965
1507 124.55
1556.9 125.116
1606.8 125.664
1656.7 126.196
1706.6 126.711
1756.5 127.212
1806.4 127.698
1856.3 128.172
1906.2 128.633
1956.1 129.081
2006 129.519
2055.9 129.946
2105.8 130.362
2155.7 130.769
2205.6 131.167
2255.5 131.555
2305.4 131.936
2355.3 132.308
2405.2 132.672
2455.1 133.028
2505 133.378
2554.9 133.721
2604.8 134.057
2654.7 134.386
2704.6 134.71
2754.5 135.027
2804.4 135.339
2854.3 135.646
2904.2 135.947
2954.1 136.243
3004 136.534
3053.9 136.82
3103.8 137.101
3153.7 137.379
3203.6 137.651
3253.5 137.92
3303.4 138.184
3353.3 138.445
3403.2 138.701
3453.1 138.954
3503 139.203
3552.9 139.449
3602.8 139.691
3652.7 139.93
3702.6 140.166
3752.5 140.398
3802.4 140.628
3852.3 140.854
3902.2 141.078
3952.1 141.299
4002 141.517
4051.9 141.732
4101.8 141.945
4151.7 142.155
4201.6 142.362
4251.5 142.567
4301.4 142.77
4351.3 142.97
4401.2 143.168
4451.1 143.364
4501 143.558
4550.9 143.75
4600.8 143.939
4650.7 144.126
4700.6 144.312
4750.5 144.495
4800.4 144.677
4850.3 144.856
4900.2 145.034
4950.1 145.21
};

\addplot [semithick, dashed, black]
table {%
10 68.7354
59.9 92.2352
109.8 101.421
159.7 107.141
209.6 111.3
259.5 114.57
309.4 117.264
359.3 119.555
409.2 121.547
459.1 123.311
509 124.892
558.9 126.325
608.8 127.636
658.7 128.844
708.6 129.963
758.5 131.006
808.4 131.983
858.3 132.901
908.2 133.767
958.1 134.587
1008 135.365
1057.9 136.106
1107.8 136.813
1157.7 137.488
1207.6 138.135
1257.5 138.756
1307.4 139.352
1357.3 139.926
1407.2 140.48
1457.1 141.014
1507 141.53
1556.9 142.03
1606.8 142.513
1656.7 142.982
1706.6 143.437
1756.5 143.879
1806.4 144.308
1856.3 144.726
1906.2 145.133
1956.1 145.529
2006 145.915
2055.9 146.292
2105.8 146.659
2155.7 147.019
2205.6 147.369
2255.5 147.712
2305.4 148.048
2355.3 148.376
2405.2 148.697
2455.1 149.012
2505 149.321
2554.9 149.623
2604.8 149.92
2654.7 150.211
2704.6 150.496
2754.5 150.776
2804.4 151.052
2854.3 151.322
2904.2 151.588
2954.1 151.849
3004 152.106
3053.9 152.358
3103.8 152.607
3153.7 152.851
3203.6 153.092
3253.5 153.329
3303.4 153.562
3353.3 153.792
3403.2 154.018
3453.1 154.242
3503 154.462
3552.9 154.678
3602.8 154.892
3652.7 155.103
3702.6 155.311
3752.5 155.516
3802.4 155.719
3852.3 155.919
3902.2 156.116
3952.1 156.311
4002 156.503
4051.9 156.693
4101.8 156.881
4151.7 157.066
4201.6 157.249
4251.5 157.43
4301.4 157.609
4351.3 157.786
4401.2 157.961
4451.1 158.134
4501 158.305
4550.9 158.474
4600.8 158.641
4650.7 158.806
4700.6 158.97
4750.5 159.132
4800.4 159.292
4850.3 159.45
4900.2 159.607
4950.1 159.763
};

\end{axis}

\end{tikzpicture}
  \caption{UMi Street Canyon}
\end{subfigure}%
\hfill%
\begin{subfigure}{0.245\textwidth}
	\centering
  \setlength\abovecaptionskip{0.1cm}
  \setlength\belowcaptionskip{-0.33cm}
	\setlength\fwidth{1.1\columnwidth}
	\setlength\fheight{0.8\columnwidth}
\begin{tikzpicture}

\definecolor{color0}{rgb}{0.12156862745098,0.466666666666667,0.705882352941177}
\definecolor{color1}{rgb}{1,0.498039215686275,0.0549019607843137}
\pgfplotsset{every tick label/.append style={font=\scriptsize}}

\begin{axis}[
width=0.951\fwidth,
height=\fheight,
at={(0\fwidth,0\fheight)},
legend cell align={left},
legend style={at={(0.03,0.97)}, anchor=north west, draw=white!80.0!black},
tick align=inside,
tick pos=left,
x grid style={white!69.01960784313725!black},
xlabel={Distance [m]},
xmin=0, xmax=148,
xtick style={color=black},
y grid style={white!69.01960784313725!black},
ymin=35, ymax=112.78033725,
ytick style={color=black},
ylabel style={font=\footnotesize\color{white!15!black}},
xlabel style={font=\footnotesize\color{white!15!black}},
ymajorgrids,
xmajorgrids,
]
\addplot [semithick, color0, mark repeat=15, mark=*, mark size={1.5}]
table {%
1 43.277197
2.49 46.998057
3.98 50.082412
5.47 52.092892
6.96 53.759981
8.45 55.053656
9.94 56.174932
11.43 56.98963
12.92 57.811834
14.41 58.568089
15.9 59.126939
17.39 60.070767
18.88 60.794616
20.37 61.576371
21.86 62.216745
23.35 62.814415
24.84 63.533844
26.33 64.000991
27.82 64.316327
29.31 64.678803
30.8 65.087904
32.29 65.278376
33.78 65.725358
35.27 65.929435
36.76 66.131586
38.25 66.458208
39.74 66.784214
41.23 66.902144
42.72 67.319563
44.21 67.790206
45.7 68.069519
47.19 68.059975
48.68 68.272322
50.17 68.64338
51.66 68.947273
53.15 69.189157
54.64 69.523235
56.13 69.770302
57.62 70.052668
59.11 70.130869
60.6 69.972068
62.09 70.116054
63.58 70.37949
65.07 70.557295
66.56 70.727272
68.05 70.814163
69.54 70.75416
71.03 70.953525
72.52 71.23242
74.01 71.49329
75.5 71.673117
76.99 72.249426
78.48 72.288908
79.97 72.355617
81.46 72.64142
82.95 72.786905
84.44 72.621548
85.93 72.607088
87.42 72.931254
88.91 73.045589
90.4 73.174261
91.89 73.164612
93.38 73.543702
94.87 73.475658
96.36 73.662067
97.85 73.724864
99.34 74.067605
100.83 74.002359
102.32 73.836799
103.81 73.640655
105.3 73.483389
106.79 73.816812
108.28 74.060365
109.77 74.209473
111.26 74.454654
112.75 74.544958
114.24 74.341782
115.73 74.371868
117.22 74.33076
118.71 74.256073
120.2 74.342408
121.69 74.329498
123.18 74.517995
124.67 74.734095
126.16 74.811223
127.65 74.854318
129.14 75.056124
130.63 75.119201
132.12 75.253248
133.61 75.225123
135.1 75.57035
136.59 75.528983
138.08 75.71848
139.57 76.09166
141.06 76.303864
142.55 76.222281
144.04 76.305516
145.53 76.282044
147.02 76.393606
148.51 76.533505
};
\addplot [semithick, color1, mark repeat=15, mark=triangle*]
table {%
1 43.162413
2.49 46.787463
3.98 50.230091
5.47 54.503779
6.96 58.2353
8.45 61.277619
9.94 63.456658
11.43 65.985582
12.92 68.733327
14.41 70.20941
15.9 71.485835
17.39 74.107637
18.88 75.567228
20.37 76.83287
21.86 76.825504
23.35 78.498855
24.84 78.665909
26.33 78.742088
27.82 79.560507
29.31 80.026078
30.8 81.882691
32.29 83.112506
33.78 83.302466
35.27 83.18624
36.76 84.434568
38.25 84.530989
39.74 85.624232
41.23 86.673897
42.72 87.301296
44.21 87.523506
45.7 89.423186
47.19 89.446572
48.68 89.951786
50.17 90.511592
51.66 91.09042
53.15 92.558239
54.64 92.475954
56.13 93.228273
57.62 93.520943
59.11 94.370151
60.6 94.378399
62.09 94.979235
63.58 95.092162
65.07 95.295791
66.56 95.540524
68.05 95.676354
69.54 95.961889
71.03 96.10842
72.52 95.905304
74.01 96.858299
75.5 97.002733
76.99 97.457978
78.48 96.836093
79.97 97.441902
81.46 97.953609
82.95 98.818859
84.44 99.478789
85.93 99.859468
87.42 100.693116
88.91 100.819919
90.4 101.199728
91.89 101.35405
93.38 101.408392
94.87 100.908162
96.36 101.406786
97.85 102.384871
99.34 101.928332
100.83 101.960122
102.32 102.081389
103.81 102.388278
105.3 102.419304
106.79 102.526309
108.28 102.514364
109.77 102.052794
111.26 102.668458
112.75 103.401761
114.24 104.683793
115.73 105.018407
117.22 104.901104
118.71 105.598523
120.2 106.494229
121.69 106.398653
123.18 106.517347
124.67 106.129902
126.16 105.997365
127.65 106.092722
129.14 106.359991
130.63 106.737587
132.12 106.539587
133.61 107.203777
135.1 106.664165
136.59 107.448362
138.08 107.416538
139.57 107.586582
141.06 107.754051
142.55 107.517729
144.04 108.235494
145.53 108.57225
147.02 109.465198
148.51 109.056253
};

\addplot [semithick, dashed, black]
table {%
1 43.0239
2.49 46.8333
3.98 49.7634
5.47 51.9527
6.96 53.6733
8.45 55.0835
9.94 56.2758
11.43 57.3073
12.92 58.2159
14.41 59.0274
15.9 59.7604
17.39 60.4286
18.88 61.0426
20.37 61.6104
21.86 62.1385
23.35 62.632
24.84 63.0952
26.33 63.5316
27.82 63.9441
29.31 64.3351
30.8 64.7069
32.29 65.0611
33.78 65.3994
35.27 65.7232
36.76 66.0336
38.25 66.3317
39.74 66.6185
41.23 66.8947
42.72 67.1611
44.21 67.4184
45.7 67.6673
47.19 67.9081
48.68 68.1414
50.17 68.3678
51.66 68.5875
53.15 68.801
54.64 69.0086
56.13 69.2106
57.62 69.4073
59.11 69.599
60.6 69.786
62.09 69.9684
63.58 70.1464
65.07 70.3204
66.56 70.4904
68.05 70.6567
69.54 70.8194
71.03 70.9786
72.52 71.1345
74.01 71.2872
75.5 71.437
76.99 71.5837
78.48 71.7277
79.97 71.869
81.46 72.0076
82.95 72.1438
84.44 72.2775
85.93 72.4089
87.42 72.538
88.91 72.665
90.4 72.7898
91.89 72.9126
93.38 73.0334
94.87 73.1523
96.36 73.2694
97.85 73.3846
99.34 73.4982
100.83 73.61
102.32 73.7202
103.81 73.8288
105.3 73.9359
106.79 74.0414
108.28 74.1455
109.77 74.2482
111.26 74.3494
112.75 74.4494
114.24 74.548
115.73 74.6453
117.22 74.7414
118.71 74.8363
120.2 74.93
121.69 75.0226
123.18 75.114
124.67 75.2043
126.16 75.2936
127.65 75.3818
129.14 75.469
130.63 75.5552
132.12 75.6404
133.61 75.7246
135.1 75.8079
136.59 75.8903
138.08 75.9718
139.57 76.0525
141.06 76.1322
142.55 76.2112
144.04 76.2893
145.53 76.3666
147.02 76.4431
148.51 76.5189
};

\addplot [semithick, dashed, black]
table {%
1 43.0239
2.49 46.8333
3.98 49.7634
5.47 54.2438
6.96 58.053
8.45 61.1751
9.94 63.8145
11.43 66.0983
12.92 68.1097
14.41 69.9062
15.9 71.529
17.39 73.0084
18.88 74.3677
20.37 75.6248
21.86 76.7939
23.35 77.8865
24.84 78.9119
26.33 79.878
27.82 80.7912
29.31 81.6569
30.8 82.4799
32.29 83.2642
33.78 84.0132
35.27 84.73
36.76 85.4172
38.25 86.0772
39.74 86.712
41.23 87.3235
42.72 87.9133
44.21 88.483
45.7 89.0338
47.19 89.567
48.68 90.0836
50.17 90.5847
51.66 91.0712
53.15 91.5438
54.64 92.0034
56.13 92.4506
57.62 92.8861
59.11 93.3105
60.6 93.7244
62.09 94.1282
63.58 94.5224
65.07 94.9076
66.56 95.284
68.05 95.6521
69.54 96.0122
71.03 96.3647
72.52 96.7099
74.01 97.048
75.5 97.3795
76.99 97.7044
78.48 98.0232
79.97 98.3359
81.46 98.6429
82.95 98.9443
84.44 99.2403
85.93 99.5312
87.42 99.8171
88.91 100.098
90.4 100.374
91.89 100.646
93.38 100.914
94.87 101.177
96.36 101.436
97.85 101.691
99.34 101.943
100.83 102.19
102.32 102.434
103.81 102.675
105.3 102.912
106.79 103.145
108.28 103.376
109.77 103.603
111.26 103.827
112.75 104.049
114.24 104.267
115.73 104.482
117.22 104.695
118.71 104.905
120.2 105.113
121.69 105.318
123.18 105.52
124.67 105.72
126.16 105.918
127.65 106.113
129.14 106.306
130.63 106.497
132.12 106.685
133.61 106.872
135.1 107.056
136.59 107.239
138.08 107.419
139.57 107.598
141.06 107.774
142.55 107.949
144.04 108.122
145.53 108.293
147.02 108.463
148.51 108.63
};

\end{axis}

\end{tikzpicture}
  \caption{Indoor Office (Mixed)}
\end{subfigure}
\caption{Average Propagation Loss vs Distance between the Nodes}
\label{fig:loss}
\end{figure*}
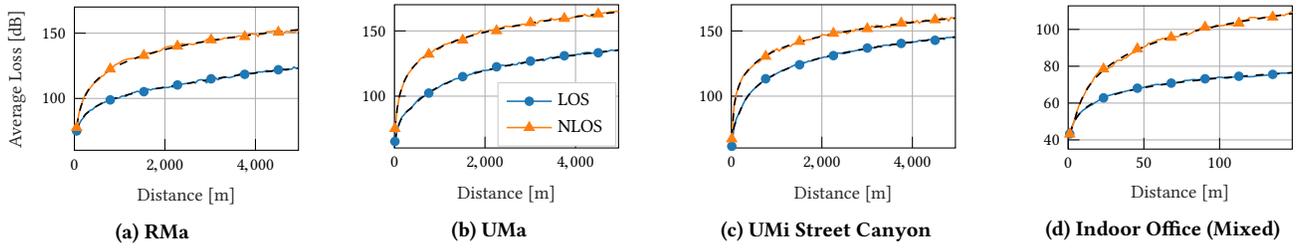

\vspace{-0.2cm}
\subsection{Examples}
The example \texttt{three-gpp-channel-example.cc} included in the \texttt{spectrum} module demonstrates the usage of the proposed framework.
It involves two devices, a transmitter and a receiver, placed at a certain distance from each other and communicating over a wireless channel.
At regular intervals, we simulate a transmission between the two nodes and estimate the \gls{snr} perceived at the receiver node.
The script provides the possibility to configure the distance between the two nodes, the channel model parameters, as well as the transmission power and the receiver noise figure. Also, it produces an output trace containing the experienced propagation loss and the \gls{snr} estimate.
As an example, in Figure~\ref{fig:loss} we reported the average propagation loss (solid lines) over the distance between the two devices operating at $2.1$~GHz for different scenarios and channel conditions, obtained by averaging the results of $100$ independent runs of this script. The black dashed lines represent the pathloss value computed from the models defined in 3GPP TR 38.901.

In the following, we review the procedure used to create and configure the channel model classes, assuming that the UMa scenario is selected:
\begin{enumerate}
  \item create an instance of the class \texttt{ThreeGppUmaChannel\-Condi\-tion\-Model};
  \item create an instance of \texttt{ThreeGppUmaPropagationLossModel}, configure the carrier frequency through the attribute "\texttt{Fre\-quency}", and set the channel condition model through the attribute "\texttt{ChannelConditionModel}";
  \item create an instance of \texttt{ThreeGppSpectrumPropagationLossModel},  set the UMa scenario, the carrier frequency and the channel condition model through the method \texttt{SetChannelModelAttribute} and using the attributes "\texttt{Scenario}", "\texttt{Frequency}" and "\texttt{ChannelConditionModel}";
  \item for each device, create an instance of   \texttt{ThreeGppAntennaArrayModel} and inform the \texttt{ThreeGppSpectrumPropagationLossModel} class about the device-antenna associations by calling the method \texttt{ThreeGppSpectrumPropagationLossModel::AddDevice}.
\end{enumerate}
Besides being implemented in the \texttt{three-gpp-channel-examp\-le.cc} script, these steps could be included in a helper class for an ns-3 module that aims at using this channel model.
For a proper usage of this model, users may need to set the transmission time granularity of the simulation based on the channel coherence time of the scenario of interest and use an error model that accounts for the non-\gls{awgn} behavior of fast fading channels.
For example, the error model in~\cite{l2sm-nr}, which has been developed according to TR 38.901, could be used in combination with the proposed channel model for \gls{nr} system-level simulations, provided that the channel coherence time is larger than the slot length of the \gls{nr} frame structure.

\vspace{-0.2cm}
\subsection{Use Cases}
The main target of the developed model is to enable system-level simulations of 3GPP scenarios through a 3GPP-compliant channel and antenna model. As such, it is a requirement for any \textbf{3GPP LTE and NR}-based system-level simulation that aims to properly model and evaluate the performance of physical layer techniques using appropriate channel modeling, both in the sub-6 GHz bands and in \gls{mmwave} bands.

Moreover, it enables the simulation and coexistence studies of different technologies that share the spectrum resources, such as 3GPP and IEEE \glspl{rat} in unlicensed/dedicated spectrum bands. For example, it can be used to evaluate the \textbf{3GPP and IEEE \gls{rat}s coexistence} of:
\begin{itemize}
    \item IEEE 802.11b/g/a/n/ac/ax (Wi-Fi) and 3GPP LTE-LAA (Licen-sed-Assisted Access) in unlicensed sub-6~GHz bands~\cite{Bojovic2019};
    \item IEEE 802.11b/g/a/n/ac/ax (Wi-Fi) and 3GPP NR-U in unlicensed sub-6 GHz bands~\cite{8891514, 9023470};
    \item IEEE 802.11ad/ay (WiGig, directional multi-Gigabit Wi-Fi) and 3GPP NR-U in unlicensed 60~GHz bands~\cite{natMag,lagen:19};
    \item IEEE 802.11p/bd, 3GPP C-V2X (Cellular V2X) and 3GPP NR V2X (Vehicle-to-Everything) in dedicated sub-6~GHz bands~\cite{8926330};
    \item IEEE 802.11bd and 3GPP NR V2X in dedicated mmWave bands~\cite{8723326};
    \item Wi-Fi, IEEE 802.11p/bd, 3GPP C-V2X and 3GPP NR V2X in unlicensed sub-6~GHz bands~\cite{8926330};
    \item WiGig, IEEE 802.11bd and 3GPP NR V2X in unlicensed 60 GHz bands~\cite{8723326}.
\end{itemize}

Also, the proposed model provides a common framework for simulations of spectrum sharing solutions through either \textbf{spectrum refarming or dynamic spectrum sharing}, for example, if different 3GPP RATs of the same operator share the licensed spectrum for some long period of time until one of the RATs becomes obsolete (spectrum refarming). This happens in low frequency bands (e.g., 900, 1800 MHz) that are essential for 3GPP NR to achieve coverage, but in which 3GPP LTE is already deployed and operational, and cannot thus be migrated to other frequency bands. As such, a key example of spectrum refarming is that of 3GPP LTE and 3GPP NR in licensed sub-6 GHz bands. Another example of spectrum sharing is
when different operators share the spectrum by means of coordination policies. In this regard, the research community has also  recently proposed solutions based on spectrum sharing~\cite{7937896} and spectrum pooling~\cite{boccardi:16} for \gls{mmwave} bands, which exploit coordination among different cellular network operators to improve the spatial reuse, and which could be tested from an end-to-end perspective on top of the proposed framework.

In addition, the developed model is also useful to evaluate the \textbf{3GPP and IEEE interworking} through a common channel modeling. 3GPP and IEEE interworking considers core network and radio access network integration by means of aggregating 3GPP-based RATs in licensed bands and Wi-Fi in unlicensed bands. Examples for which the developed model could be used include:
\begin{itemize}
    \item Wi-Fi and 3GPP LTE interworking, e.g., through LTE-WLAN Aggregation (LWA) and LTE-WLAN Radio Level Integration with IPsec Tunnel (LWIP)~\cite{lteAgg},
    \item IEEE 802.11ax and 3GPP NR interworking~\cite{ranConv}.
\end{itemize}

\vspace{-0.2cm}
\section{Conclusions and Future Work}
\label{sec:conclusions}
In this paper, we introduced a new channel model for ns-3, developed following the specifications in \cite{TR38901}. This work is expected to enrich ns-3 by enabling a more accurate modeling of the dynamics of wireless channels between $0.5$ and $100$~GHz, thus enhancing the support for the simulation of wireless systems.
In Section~\ref{sec:intro}, we introduced the importance of channel models and simulation tools for the design of next-generation wireless systems. In Section~\ref{sec:channel}, we explained the motivations that drive the design of more accurate channel models and described the \glspl{scm} approach.
In Section~\ref{sec:implementation}, we overviewed the 3GPP \gls{scm} and explained how it has been implemented in ns-3. Finally, in  Section~\ref{sec:scenarios}, we provided an example of usage and outlined some possible use cases.

We plan to further improve this work by (i) refining the antenna model to enable the modeling of multiple panels with dual polarization; (ii) implementing the outdoor-to-indoor penetration loss model described in \cite{TR38901}, Section~7.4.3; (iii) implementing the additional modeling components, such as the spatial consistency procedure and the modeling of the oxygen absorption, specified in \cite{TR38901}, Section~7.6; (iv) performing a calibration campaign to validate the model following the assumptions reported in \cite{TR38901}, Section~7.8; and (v) lowering the computation time needed to generate a channel realization~\cite{testolina2020scalable} through the optimization of matrix operations.
Moreover, thanks to the modular design, this work can be easily extended to support other \gls{scm}-based channel models, hence providing a common framework for the simulation of wireless systems.

\vspace{-.1cm}
\begin{acks}
The work of Tommaso Zugno was partially funded by the Google Summer of Code 2019 program. Also, the work of the authors from the University of Padova has been partially supported by NIST through Awards No. 70NANB17H166 and 70NANB18H273, and CTTC authors have received funding from the Spanish MINECO grant TEC2017-88373-R (5G-REFINE) and Generalitat de Catalunya grant 2017 SGR 1195. Finally, the authors would like to thank Tom Henderson for his useful suggestions and support during this work.
\end{acks}

\vspace{-0.15cm}
\balance
\bibliographystyle{ACM-Reference-Format.bst}
\bibliography{./bibl.bib}


\newcommand{\SortNoop}[1]{}
\begin{thebibliography}{42}


\ifx \showCODEN    \undefined \def \showCODEN     #1{\unskip}     \fi
\ifx \showDOI      \undefined \def \showDOI       #1{#1}\fi
\ifx \showISBNx    \undefined \def \showISBNx     #1{\unskip}     \fi
\ifx \showISBNxiii \undefined \def \showISBNxiii  #1{\unskip}     \fi
\ifx \showISSN     \undefined \def \showISSN      #1{\unskip}     \fi
\ifx \showLCCN     \undefined \def \showLCCN      #1{\unskip}     \fi
\ifx \shownote     \undefined \def \shownote      #1{#1}          \fi
\ifx \showarticletitle \undefined \def \showarticletitle #1{#1}   \fi
\ifx \showURL      \undefined \def \showURL       {\relax}        \fi
\providecommand\bibfield[2]{#2}
\providecommand\bibinfo[2]{#2}
\providecommand\natexlab[1]{#1}
\providecommand\showeprint[2][]{arXiv:#2}

\bibitem[\protect\citeauthoryear{{3GPP}}{{3GPP}}{2017a}]%
        {38900}
\bibfield{author}{\bibinfo{person}{{3GPP}}.} \bibinfo{year}{2017}\natexlab{a}.
\newblock \bibinfo{title}{{Study on Channel Model for Frequency Spectrum Above
  6 GHz}}.
\newblock \bibinfo{howpublished}{TR 38.900 (Rel. 14)}.
  (\bibinfo{year}{2017}).
\newblock


\bibitem[\protect\citeauthoryear{{3GPP}}{{3GPP}}{2017b}]%
        {3GPP38913}
\bibfield{author}{\bibinfo{person}{{3GPP}}.} \bibinfo{year}{2017}\natexlab{b}.
\newblock \bibinfo{title}{{TR 38.913, Study on Scenarios and Requirements for
  Next Generation Access Technologies, V14.1.0}}.
\newblock   (\bibinfo{year}{2017}).
\newblock


\bibitem[\protect\citeauthoryear{3GPP}{3GPP}{2018}]%
        {38300}
\bibfield{author}{\bibinfo{person}{3GPP}.} \bibinfo{year}{2018}\natexlab{}.
\newblock \bibinfo{title}{{NR and NG-RAN Overall Description}}.
\newblock \bibinfo{howpublished}{TS 38.300 (Rel. 15)}.
  (\bibinfo{year}{2018}).
\newblock


\bibitem[\protect\citeauthoryear{{3GPP}}{{3GPP}}{2019}]%
        {TR38901}
\bibfield{author}{\bibinfo{person}{{3GPP}}.} \bibinfo{year}{Jun.
  2019}\natexlab{}.
\newblock \bibinfo{title}{{Study on Channel Model for Frequencies from 0.5 to
  100 GHz}}.
\newblock \bibinfo{howpublished}{TR 38.901 (Rel. 15), V15.0.0}.
  (\bibinfo{year}{Jun. 2019}).
\newblock


\bibitem[\protect\citeauthoryear{{Andrews}, {Bai}, {Kulkarni}, {Alkhateeb},
  {Gupta}, and {Heath}}{{Andrews} et~al\mbox{.}}{2017}]%
        {andrews2017modeling}
\bibfield{author}{\bibinfo{person}{J.~G. {Andrews}}, \bibinfo{person}{T.
  {Bai}}, \bibinfo{person}{M.~N. {Kulkarni}}, \bibinfo{person}{A. {Alkhateeb}},
  \bibinfo{person}{A.~K. {Gupta}}, {and} \bibinfo{person}{R.~W. {Heath}}.}
  \bibinfo{year}{2017}\natexlab{}.
\newblock \showarticletitle{Modeling and Analyzing Millimeter Wave Cellular
  Systems}.
\newblock \bibinfo{journal}{\emph{IEEE Transactions on Communications}}
  \bibinfo{volume}{65}, \bibinfo{number}{1} (\bibinfo{date}{Jan}
  \bibinfo{year}{2017}), pp. \bibinfo{pages}{403--430}.
\newblock
\showISSN{0090-6778}


\bibitem[\protect\citeauthoryear{Assasa and Widmer}{Assasa and Widmer}{2017}]%
        {assasa2017extending}
\bibfield{author}{\bibinfo{person}{H. Assasa} {and} \bibinfo{person}{J.
  Widmer}.} \bibinfo{year}{2017}\natexlab{}.
\newblock \showarticletitle{{Extending the IEEE 802.11 ad Model: Scheduled
  Access, Spatial Reuse, Clustering, and Relaying}}. In
  \bibinfo{booktitle}{\emph{Proceedings of the Workshop on ns-3}}
  \emph{(\bibinfo{series}{WNS3 '17})}. \bibinfo{publisher}{ACM},
  \bibinfo{address}{Porto, Portugal}, pp. \bibinfo{pages}{39--46}.
\newblock


\bibitem[\protect\citeauthoryear{Assasa, Widmer, Ropitault, Bodi, and
  Golmie}{Assasa et~al\mbox{.}}{2019b}]%
        {assasa2019high}
\bibfield{author}{\bibinfo{person}{H. Assasa}, \bibinfo{person}{J. Widmer},
  \bibinfo{person}{T. Ropitault}, \bibinfo{person}{A. Bodi}, {and}
  \bibinfo{person}{N. Golmie}.} \bibinfo{year}{2019}\natexlab{b}.
\newblock \showarticletitle{{High Fidelity Simulation of IEEE 802.11ad in ns-3
  Using a Quasi-Deterministic Channel Model}}. In
  \bibinfo{booktitle}{\emph{Proceedings of the 2019 Workshop on Next-Generation
  Wireless with ns-3}} \emph{(\bibinfo{series}{WNGW 2019})}.
  \bibinfo{publisher}{ACM}, \bibinfo{address}{Florence, Italy}, pp.
  \bibinfo{pages}{22--25}.
\newblock
\showISBNx{9781450372787}


\bibitem[\protect\citeauthoryear{Assasa, Widmer, Ropitault, and Golmie}{Assasa
  et~al\mbox{.}}{2019a}]%
        {assasa11ad2019}
\bibfield{author}{\bibinfo{person}{H. Assasa}, \bibinfo{person}{J. Widmer},
  \bibinfo{person}{T. Ropitault}, {and} \bibinfo{person}{N. Golmie}.}
  \bibinfo{year}{2019}\natexlab{a}.
\newblock \showarticletitle{{Enhancing the ns-3 IEEE 802.11ad Model Fidelity:
  Beam Codebooks, Multi-Antenna Beamforming Training, and Quasi-Deterministic
  MmWave Channel}}. In \bibinfo{booktitle}{\emph{{Proceedings of the 2019
  Workshop on ns-3}}} \emph{(\bibinfo{series}{WNS3 2019})}.
  \bibinfo{publisher}{ACM}, \bibinfo{address}{Florence, Italy}, pp.
  \bibinfo{pages}{33--40}.
\newblock


\bibitem[\protect\citeauthoryear{Baldo and Miozzo}{Baldo and Miozzo}{2009}]%
        {baldo2009spectrum}
\bibfield{author}{\bibinfo{person}{N. Baldo} {and} \bibinfo{person}{M.
  Miozzo}.} \bibinfo{year}{2009}\natexlab{}.
\newblock \showarticletitle{{Spectrum-Aware Channel and PHY Layer Modeling for
  ns-3}}. In \bibinfo{booktitle}{\emph{Proceedings of the Fourth International
  ICST Conference on Performance Evaluation Methodologies and Tools}}
  \emph{(\bibinfo{series}{VALUETOOLS 2009})}. ICST, \bibinfo{address}{Pisa,
  Italy}.
\newblock


\bibitem[\protect\citeauthoryear{Boccardi, {Heath Jr}, Lozano, Marzetta, and
  Popovski}{Boccardi et~al\mbox{.}}{2014}]%
        {BocHLMP:14}
\bibfield{author}{\bibinfo{person}{F. Boccardi}, \bibinfo{person}{R.~W. {Heath
  Jr}}, \bibinfo{person}{A. Lozano}, \bibinfo{person}{T.~L. Marzetta}, {and}
  \bibinfo{person}{P. Popovski}.} \bibinfo{year}{2014}\natexlab{}.
\newblock \showarticletitle{Five Disruptive Technology Directions for {5G}}.
\newblock \bibinfo{journal}{\emph{IEEE Communication Magazine}}
  \bibinfo{volume}{52}, \bibinfo{number}{2} (\bibinfo{date}{Feb.}
  \bibinfo{year}{2014}), pp. \bibinfo{pages}{74--80}.
\newblock


\bibitem[\protect\citeauthoryear{{Boccardi}, {Shokri-Ghadikolaei}, {Fodor},
  {Erkip}, {Fischione}, {Kountouris}, {Popovski}, and {Zorzi}}{{Boccardi}
  et~al\mbox{.}}{2016}]%
        {boccardi:16}
\bibfield{author}{\bibinfo{person}{F. {Boccardi}}, \bibinfo{person}{H.
  {Shokri-Ghadikolaei}}, \bibinfo{person}{G. {Fodor}}, \bibinfo{person}{E.
  {Erkip}}, \bibinfo{person}{C. {Fischione}}, \bibinfo{person}{M.
  {Kountouris}}, \bibinfo{person}{P. {Popovski}}, {and} \bibinfo{person}{M.
  {Zorzi}}.} \bibinfo{year}{2016}\natexlab{}.
\newblock \showarticletitle{Spectrum Pooling in MmWave Networks: Opportunities,
  Challenges, and Enablers}.
\newblock \bibinfo{journal}{\emph{IEEE Communications Magazine}}
  \bibinfo{volume}{54}, \bibinfo{number}{11} (\bibinfo{date}{November}
  \bibinfo{year}{2016}), pp. \bibinfo{pages}{33--39}.
\newblock


\bibitem[\protect\citeauthoryear{{Bojovic}, {Giupponi}, {Ali}, and
  {Miozzo}}{{Bojovic} et~al\mbox{.}}{2019}]%
        {Bojovic2019}
\bibfield{author}{\bibinfo{person}{B. {Bojovic}}, \bibinfo{person}{L.
  {Giupponi}}, \bibinfo{person}{Z. {Ali}}, {and} \bibinfo{person}{M.
  {Miozzo}}.} \bibinfo{year}{2019}\natexlab{}.
\newblock \showarticletitle{Evaluating Unlicensed {LTE} Technologies: {LAA} vs
  {LTE-U}}.
\newblock \bibinfo{journal}{\emph{IEEE Access}}  \bibinfo{volume}{7}
  (\bibinfo{year}{2019}), pp. \bibinfo{pages}{89714--89751}.
\newblock


\bibitem[\protect\citeauthoryear{{Chen}, {Xu}, and {Jiang}}{{Chen}
  et~al\mbox{.}}{2019}]%
        {8891514}
\bibfield{author}{\bibinfo{person}{Q. {Chen}}, \bibinfo{person}{X. {Xu}}, {and}
  \bibinfo{person}{H. {Jiang}}.} \bibinfo{year}{2019}\natexlab{}.
\newblock \showarticletitle{Spatial Multiplexing Based NR-U and WiFi
  Coexistence in Unlicensed Spectrum}. In \bibinfo{booktitle}{\emph{2019 IEEE
  90th Vehicular Technology Conference (VTC2019-Fall)}}.
  \bibinfo{address}{Honolulu, HI, USA}, pp. \bibinfo{pages}{1--5}.
\newblock


\bibitem[\protect\citeauthoryear{Ferrand, Amara, Valentin, and
  Guillaud}{Ferrand et~al\mbox{.}}{2016}]%
        {ferrand2016trends}
\bibfield{author}{\bibinfo{person}{P. Ferrand}, \bibinfo{person}{M. Amara},
  \bibinfo{person}{S. Valentin}, {and} \bibinfo{person}{M. Guillaud}.}
  \bibinfo{year}{2016}\natexlab{}.
\newblock \showarticletitle{Trends and Challenges in Wireless Channel Modeling
  for Evolving Radio Access}.
\newblock \bibinfo{journal}{\emph{IEEE Communication Magazine}}
  \bibinfo{volume}{54}, \bibinfo{number}{7} (\bibinfo{date}{July}
  \bibinfo{year}{2016}), pp. \bibinfo{pages}{93--99}.
\newblock
\showISSN{0163-6804}


\bibitem[\protect\citeauthoryear{{Ghafoor}, {Guizani}, {Kong}, {Maghdid}, and
  {Jasim}}{{Ghafoor} et~al\mbox{.}}{2019}]%
        {8926330}
\bibfield{author}{\bibinfo{person}{K.~Z. {Ghafoor}}, \bibinfo{person}{M.
  {Guizani}}, \bibinfo{person}{L. {Kong}}, \bibinfo{person}{H.~S. {Maghdid}},
  {and} \bibinfo{person}{K.~F. {Jasim}}.} \bibinfo{year}{2019}\natexlab{}.
\newblock \showarticletitle{Enabling Efficient Coexistence of DSRC and C-V2X in
  Vehicular Networks}.
\newblock \bibinfo{journal}{\emph{IEEE Wireless Communications (Early Access)}}
  (\bibinfo{year}{2019}), pp. \bibinfo{pages}{2--8}.
\newblock


\bibitem[\protect\citeauthoryear{Hemadeh, Satyanarayana, El-Hajjar, and
  Hanzo}{Hemadeh et~al\mbox{.}}{2018}]%
        {hemadeh2018millimeter}
\bibfield{author}{\bibinfo{person}{I. Hemadeh}, \bibinfo{person}{K.
  Satyanarayana}, \bibinfo{person}{M. El-Hajjar}, {and} \bibinfo{person}{L.
  Hanzo}.} \bibinfo{year}{2018}\natexlab{}.
\newblock \showarticletitle{Millimeter-Wave Communications: Physical Channel
  Models, Design Considerations, Antenna Constructions and Link-Budget}.
\newblock \bibinfo{journal}{\emph{IEEE Communications Surveys Tutorials}}
  \bibinfo{volume}{20}, \bibinfo{number}{2} (\bibinfo{year}{2018}), pp.
  \bibinfo{pages}{870 -- 913}.
\newblock


\bibitem[\protect\citeauthoryear{{Khorov}, {Kiryanov}, {Lyakhov}, and
  {Bianchi}}{{Khorov} et~al\mbox{.}}{2019}]%
        {khorov2019tutorial}
\bibfield{author}{\bibinfo{person}{E. {Khorov}}, \bibinfo{person}{A.
  {Kiryanov}}, \bibinfo{person}{A. {Lyakhov}}, {and} \bibinfo{person}{G.
  {Bianchi}}.} \bibinfo{year}{2019}\natexlab{}.
\newblock \showarticletitle{{A Tutorial on IEEE 802.11ax High Efficiency
  WLANs}}.
\newblock \bibinfo{journal}{\emph{IEEE Communications Surveys Tutorials}}
  \bibinfo{volume}{21}, \bibinfo{number}{1} (\bibinfo{date}{Firstquarter}
  \bibinfo{year}{2019}), pp. \bibinfo{pages}{197--216}.
\newblock
\showISSN{2373-745X}


\bibitem[\protect\citeauthoryear{{Koymen}, {Partyka}, {Subramanian}, and
  {Li}}{{Koymen} et~al\mbox{.}}{2015}]%
        {koymen2015indoor}
\bibfield{author}{\bibinfo{person}{O.~H. {Koymen}}, \bibinfo{person}{A.
  {Partyka}}, \bibinfo{person}{S. {Subramanian}}, {and} \bibinfo{person}{J.
  {Li}}.} \bibinfo{year}{2015}\natexlab{}.
\newblock \showarticletitle{{Indoor mm-Wave Channel Measurements: Comparative
  Study of 2.9 GHz and 29 GHz}}. In \bibinfo{booktitle}{\emph{IEEE Global
  Communications Conference (GLOBECOM)}}. \bibinfo{address}{San Diego, CA,
  USA}, pp. \bibinfo{pages}{1--6}.
\newblock
\showISSN{null}


\bibitem[\protect\citeauthoryear{{Lagen}, {Giupponi}, {Goyal}, {Patriciello},
  {Bojovic}, {Demir}, and {Beluri}}{{Lagen} et~al\mbox{.}}{2020}]%
        {lagen:19}
\bibfield{author}{\bibinfo{person}{S. {Lagen}}, \bibinfo{person}{L.
  {Giupponi}}, \bibinfo{person}{S. {Goyal}}, \bibinfo{person}{N.
  {Patriciello}}, \bibinfo{person}{B. {Bojovic}}, \bibinfo{person}{A. {Demir}},
  {and} \bibinfo{person}{M. {Beluri}}.} \bibinfo{year}{2020}\natexlab{}.
\newblock \showarticletitle{New Radio Beam-based Access to Unlicensed Spectrum:
  Design Challenges and Solutions}.
\newblock \bibinfo{journal}{\emph{IEEE Communications Surveys Tutorials}}
  \bibinfo{volume}{22}, \bibinfo{number}{1} (\bibinfo{date}{Mar.}
  \bibinfo{year}{2020}), pp. \bibinfo{pages}{8--37}.
\newblock


\bibitem[\protect\citeauthoryear{Lagen, Wanuga, Elkotby, Goyal, Patriciello,
  and Giupponi}{Lagen et~al\mbox{.}}{2020}]%
        {l2sm-nr}
\bibfield{author}{\bibinfo{person}{S. Lagen}, \bibinfo{person}{K. Wanuga},
  \bibinfo{person}{H. Elkotby}, \bibinfo{person}{S. Goyal}, \bibinfo{person}{N.
  Patriciello}, {and} \bibinfo{person}{L. Giupponi}.}
  \bibinfo{year}{2020}\natexlab{}.
\newblock \showarticletitle{{New Radio Physical Layer Abstraction for
  System-Level Simulations of 5G Networks}}. In
  \bibinfo{booktitle}{\emph{Proceedings of IEEE International Conference on
  Communications}} \emph{(\bibinfo{series}{IEEE ICC})}.
  \bibinfo{address}{Dublin, Ireland}.
\newblock


\bibitem[\protect\citeauthoryear{Lanante, Roy, Carpenter, and Deronne}{Lanante
  et~al\mbox{.}}{2019}]%
        {lanante2019improved}
\bibfield{author}{\bibinfo{person}{L. Lanante}, \bibinfo{person}{S. Roy},
  \bibinfo{person}{S.~E. Carpenter}, {and} \bibinfo{person}{S. Deronne}.}
  \bibinfo{year}{2019}\natexlab{}.
\newblock \showarticletitle{{Improved Abstraction for Clear Channel Assessment
  in ns-3 802.11 WLAN Model}}. In \bibinfo{booktitle}{\emph{Proceedings of the
  2019 Workshop on ns-3}} \emph{(\bibinfo{series}{WNS3 2019})}.
  \bibinfo{address}{Florence, Italy}, pp. \bibinfo{pages}{49--56}.
\newblock
\showISBNx{9781450371407}


\bibitem[\protect\citeauthoryear{Lecci, Testolina, Giordani, Polese, Ropitault,
  Gentile, Varshney, Bodi, and Zorzi}{Lecci et~al\mbox{.}}{2020}]%
        {lecci2020simplified}
\bibfield{author}{\bibinfo{person}{M. Lecci}, \bibinfo{person}{P. Testolina},
  \bibinfo{person}{M. Giordani}, \bibinfo{person}{M. Polese},
  \bibinfo{person}{T. Ropitault}, \bibinfo{person}{C. Gentile},
  \bibinfo{person}{N. Varshney}, \bibinfo{person}{A. Bodi}, {and}
  \bibinfo{person}{M. Zorzi}.} \bibinfo{year}{2020}\natexlab{}.
\newblock \showarticletitle{{Simplified Ray Tracing for the Millimeter Wave
  Channel: A Performance Evaluation}}. In \bibinfo{booktitle}{\emph{Information
  Theory and Applications Workshop (ITA)}}. \bibinfo{address}{San Diego, CA,
  USA}.
\newblock


\bibitem[\protect\citeauthoryear{{Maldonado}, {Rosa}, and
  {Pedersen}}{{Maldonado} et~al\mbox{.}}{2020}]%
        {9023470}
\bibfield{author}{\bibinfo{person}{R. {Maldonado}}, \bibinfo{person}{C.
  {Rosa}}, {and} \bibinfo{person}{K.~I. {Pedersen}}.}
  \bibinfo{year}{2020}\natexlab{}.
\newblock \showarticletitle{Latency and Reliability Analysis of Cellular
  Networks in Unlicensed Spectrum}.
\newblock \bibinfo{journal}{\emph{IEEE Access}}  \bibinfo{volume}{8}
  (\bibinfo{year}{2020}), pp. \bibinfo{pages}{49412--49423}.
\newblock


\bibitem[\protect\citeauthoryear{{Maltsev}, {Pudeyev}, {Lomayev}, and
  {Bolotin}}{{Maltsev} et~al\mbox{.}}{2016}]%
        {maltsev2016channel}
\bibfield{author}{\bibinfo{person}{A. {Maltsev}}, \bibinfo{person}{A.
  {Pudeyev}}, \bibinfo{person}{A. {Lomayev}}, {and} \bibinfo{person}{I.
  {Bolotin}}.} \bibinfo{year}{2016}\natexlab{}.
\newblock \showarticletitle{{Channel Modeling in the Next Generation mmWave
  Wi-Fi: IEEE 802.11ay Standard}}. In \bibinfo{booktitle}{\emph{22th European
  Wireless Conference}}. \bibinfo{address}{Oulu, Finland}.
\newblock


\bibitem[\protect\citeauthoryear{{Mezzavilla}, {Zhang}, {Polese}, {Ford},
  {Dutta}, {Rangan}, and {Zorzi}}{{Mezzavilla} et~al\mbox{.}}{2018}]%
        {mezzavilla2017end}
\bibfield{author}{\bibinfo{person}{M. {Mezzavilla}}, \bibinfo{person}{M.
  {Zhang}}, \bibinfo{person}{M. {Polese}}, \bibinfo{person}{R. {Ford}},
  \bibinfo{person}{S. {Dutta}}, \bibinfo{person}{S. {Rangan}}, {and}
  \bibinfo{person}{M. {Zorzi}}.} \bibinfo{year}{2018}\natexlab{}.
\newblock \showarticletitle{{End-to-End Simulation of 5G mmWave Networks}}.
\newblock \bibinfo{journal}{\emph{IEEE Communications Surveys \& Tutorials}}
  \bibinfo{volume}{20}, \bibinfo{number}{3} (\bibinfo{date}{Third quarter}
  \bibinfo{year}{2018}), pp. \bibinfo{pages}{2237--2263}.
\newblock
\showISSN{1553-877X}


\bibitem[\protect\citeauthoryear{{N. Patriciello, S. Goyal, S. Lagen, L.
  Giupponi, B. Bojovic, A. Demir, M. Beluri}}{{N. Patriciello, S. Goyal, S.
  Lagen, L. Giupponi, B. Bojovic, A. Demir, M. Beluri}}{2019}]%
        {natMag}
\bibfield{author}{\bibinfo{person}{{N. Patriciello, S. Goyal, S. Lagen, L.
  Giupponi, B. Bojovic, A. Demir, M. Beluri}}.}
  \bibinfo{year}{2019}\natexlab{}.
\newblock \bibinfo{title}{{NR-U and WiGig Coexistence in 60 GHz Bands}}.
\newblock \bibinfo{howpublished}{arXiv preprint arXiv:2001.04779}.
  (\bibinfo{year}{2019}).
\newblock


\bibitem[\protect\citeauthoryear{{Naik}, {Choudhury}, and {Park}}{{Naik}
  et~al\mbox{.}}{2019}]%
        {8723326}
\bibfield{author}{\bibinfo{person}{G. {Naik}}, \bibinfo{person}{B.
  {Choudhury}}, {and} \bibinfo{person}{J. {Park}}.}
  \bibinfo{year}{2019}\natexlab{}.
\newblock \showarticletitle{{IEEE 802.11bd 5G NR V2X: Evolution of Radio Access
  Technologies for V2X Communications}}.
\newblock \bibinfo{journal}{\emph{IEEE Access}}  \bibinfo{volume}{7}
  (\bibinfo{year}{2019}), pp. \bibinfo{pages}{70169--70184}.
\newblock


\bibitem[\protect\citeauthoryear{Nuggehalli}{Nuggehalli}{2016}]%
        {lteAgg}
\bibfield{author}{\bibinfo{person}{P. Nuggehalli}.}
  \bibinfo{year}{2016}\natexlab{}.
\newblock \showarticletitle{{LTE-WLAN Aggregation [Industry Perspectives]}}.
\newblock \bibinfo{journal}{\emph{IEEE Wireless Communications}}
  \bibinfo{volume}{23}, \bibinfo{number}{4} (\bibinfo{date}{Aug.}
  \bibinfo{year}{2016}), pp. \bibinfo{pages}{4--6}.
\newblock


\bibitem[\protect\citeauthoryear{Patriciello, Lagen, Bojovic, and
  Giupponi}{Patriciello et~al\mbox{.}}{2019}]%
        {PATRICIELLO2019101933}
\bibfield{author}{\bibinfo{person}{N. Patriciello}, \bibinfo{person}{S. Lagen},
  \bibinfo{person}{B. Bojovic}, {and} \bibinfo{person}{L. Giupponi}.}
  \bibinfo{year}{2019}\natexlab{}.
\newblock \showarticletitle{{An E2E Simulator for 5G NR Networks}}.
\newblock \bibinfo{journal}{\emph{Simulation Modelling Practice and Theory}}
  \bibinfo{volume}{96} (\bibinfo{year}{2019}), pp. \bibinfo{pages}{101933}.
\newblock
\showISSN{1569-190X}


\bibitem[\protect\citeauthoryear{Polese, Restuccia, Gosain, Jornet, Bhardwaj,
  Ariyarathna, Mandal, Zheng, Dhananjay, Mezzavilla, Buckwalter, Rodwell, Wang,
  Zorzi, Madanayake, and Melodia}{Polese et~al\mbox{.}}{2019}]%
        {polese2019millimetera}
\bibfield{author}{\bibinfo{person}{M. Polese}, \bibinfo{person}{F. Restuccia},
  \bibinfo{person}{A. Gosain}, \bibinfo{person}{J. Jornet}, \bibinfo{person}{S.
  Bhardwaj}, \bibinfo{person}{V. Ariyarathna}, \bibinfo{person}{S. Mandal},
  \bibinfo{person}{K. Zheng}, \bibinfo{person}{A. Dhananjay},
  \bibinfo{person}{M. Mezzavilla}, \bibinfo{person}{J. Buckwalter},
  \bibinfo{person}{M. Rodwell}, \bibinfo{person}{X. Wang}, \bibinfo{person}{M.
  Zorzi}, \bibinfo{person}{A. Madanayake}, {and} \bibinfo{person}{T. Melodia}.}
  \bibinfo{year}{2019}\natexlab{}.
\newblock \showarticletitle{{MillimeTera: Toward A Large-Scale Open-Source
  MmWave and Terahertz Experimental Testbed}}. In
  \bibinfo{booktitle}{\emph{Proceedings of the 3rd ACM Workshop on
  Millimeter-Wave Networks and Sensing Systems}} \emph{(\bibinfo{series}{mmNets
  '19})}. \bibinfo{publisher}{ACM}, \bibinfo{address}{Los Cabos, Mexico}, pp.
  \bibinfo{pages}{27--32}.
\newblock
\showISBNx{9781450369329}


\bibitem[\protect\citeauthoryear{{Polese} and {Zorzi}}{{Polese} and
  {Zorzi}}{2018}]%
        {polese2018impact}
\bibfield{author}{\bibinfo{person}{M. {Polese}} {and} \bibinfo{person}{M.
  {Zorzi}}.} \bibinfo{year}{2018}\natexlab{}.
\newblock \showarticletitle{{Impact of Channel Models on the End-to-End
  Performance of mmWave Cellular Networks}}. In \bibinfo{booktitle}{\emph{IEEE
  19th International Workshop on Signal Processing Advances in Wireless
  Communications (SPAWC)}}. \bibinfo{address}{Kalamata, Greece}.
\newblock


\bibitem[\protect\citeauthoryear{Rappaport, Sun, Mayzus, Zhao, Azar, Wang,
  Wong, Schulz, Samimi, and Gutierrez}{Rappaport et~al\mbox{.}}{2013}]%
        {rappaportmillimeter}
\bibfield{author}{\bibinfo{person}{T.~S. Rappaport}, \bibinfo{person}{S. Sun},
  \bibinfo{person}{R. Mayzus}, \bibinfo{person}{H. Zhao}, \bibinfo{person}{Y.
  Azar}, \bibinfo{person}{K. Wang}, \bibinfo{person}{G.~N. Wong},
  \bibinfo{person}{J.~K. Schulz}, \bibinfo{person}{M. Samimi}, {and}
  \bibinfo{person}{F. Gutierrez}.} \bibinfo{year}{2013}\natexlab{}.
\newblock \showarticletitle{{Millimeter Wave Mobile Communications for 5G
  Cellular: It Will Work!}}
\newblock \bibinfo{journal}{\emph{IEEE Access}}  \bibinfo{volume}{1}
  (\bibinfo{date}{May} \bibinfo{year}{2013}), pp. \bibinfo{pages}{335--349}.
\newblock


\bibitem[\protect\citeauthoryear{{Rebato}, {Boccardi}, {Mezzavilla}, {Rangan},
  and {Zorzi}}{{Rebato} et~al\mbox{.}}{2017}]%
        {7937896}
\bibfield{author}{\bibinfo{person}{M. {Rebato}}, \bibinfo{person}{F.
  {Boccardi}}, \bibinfo{person}{M. {Mezzavilla}}, \bibinfo{person}{S.
  {Rangan}}, {and} \bibinfo{person}{M. {Zorzi}}.}
  \bibinfo{year}{2017}\natexlab{}.
\newblock \showarticletitle{Hybrid Spectrum Sharing in mmWave Cellular
  Networks}.
\newblock \bibinfo{journal}{\emph{IEEE Transactions on Cognitive Communications
  and Networking}} \bibinfo{volume}{3}, \bibinfo{number}{2}
  (\bibinfo{date}{June} \bibinfo{year}{2017}), pp. \bibinfo{pages}{155--168}.
\newblock
\showISSN{2372-2045}


\bibitem[\protect\citeauthoryear{{Rebato}, {Polese}, and {Zorzi}}{{Rebato}
  et~al\mbox{.}}{2018}]%
        {rebato2018multi}
\bibfield{author}{\bibinfo{person}{M. {Rebato}}, \bibinfo{person}{M. {Polese}},
  {and} \bibinfo{person}{M. {Zorzi}}.} \bibinfo{year}{2018}\natexlab{}.
\newblock \showarticletitle{{Multi-Sector and Multi-Panel Performance in 5G
  mmWave Cellular Networks}}. In \bibinfo{booktitle}{\emph{2018 IEEE Global
  Communications Conference (GLOBECOM)}}. \bibinfo{address}{Abu Dhabi, United
  Arab Emirates}, pp. \bibinfo{pages}{1--6}.
\newblock
\showISSN{1930-529X}


\bibitem[\protect\citeauthoryear{{Remley}, {Gordon}, {Novotny}, {Curtin},
  {Holloway}, {Simons}, {Horansky}, {Allman}, {Senic}, {Becker}, {Jargon},
  {Hale}, {Williams}, {Feldman}, {Cheron}, {Chamberlin}, {Gentile}, {Senic},
  {Sun}, {Papazian}, {Quimby}, {Mujumdar}, and {Golmie}}{{Remley}
  et~al\mbox{.}}{2017}]%
        {remley2017measurement}
\bibfield{author}{\bibinfo{person}{K.~A. {Remley}}, \bibinfo{person}{J.~A.
  {Gordon}}, \bibinfo{person}{D. {Novotny}}, \bibinfo{person}{A.~E. {Curtin}},
  \bibinfo{person}{C.~L. {Holloway}}, \bibinfo{person}{M.~T. {Simons}},
  \bibinfo{person}{R.~D. {Horansky}}, \bibinfo{person}{M.~S. {Allman}},
  \bibinfo{person}{D. {Senic}}, \bibinfo{person}{M. {Becker}},
  \bibinfo{person}{J.~A. {Jargon}}, \bibinfo{person}{P.~D. {Hale}},
  \bibinfo{person}{D.~F. {Williams}}, \bibinfo{person}{A. {Feldman}},
  \bibinfo{person}{J. {Cheron}}, \bibinfo{person}{R. {Chamberlin}},
  \bibinfo{person}{C. {Gentile}}, \bibinfo{person}{J. {Senic}},
  \bibinfo{person}{R. {Sun}}, \bibinfo{person}{P.~B. {Papazian}},
  \bibinfo{person}{J. {Quimby}}, \bibinfo{person}{M. {Mujumdar}}, {and}
  \bibinfo{person}{N. {Golmie}}.} \bibinfo{year}{2017}\natexlab{}.
\newblock \showarticletitle{{Measurement Challenges for 5G and Beyond: An
  Update from the National Institute of Standards and Technology}}.
\newblock \bibinfo{journal}{\emph{IEEE Microwave Magazine}}
  \bibinfo{volume}{18}, \bibinfo{number}{5} (\bibinfo{date}{July}
  \bibinfo{year}{2017}), pp. \bibinfo{pages}{41--56}.
\newblock
\showISSN{1557-9581}


\bibitem[\protect\citeauthoryear{Saha, Ghasempour, Haider, Siddiqui, Melo,
  Somanchi, Zakrajsek, Singh, Shyamsunder, Torres, et~al\mbox{.}}{Saha
  et~al\mbox{.}}{2019}]%
        {saha2019x60}
\bibfield{author}{\bibinfo{person}{S.~K. Saha}, \bibinfo{person}{Y.
  Ghasempour}, \bibinfo{person}{M.~K. Haider}, \bibinfo{person}{T. Siddiqui},
  \bibinfo{person}{P.~De Melo}, \bibinfo{person}{N. Somanchi},
  \bibinfo{person}{L. Zakrajsek}, \bibinfo{person}{A. Singh},
  \bibinfo{person}{R. Shyamsunder}, \bibinfo{person}{O. Torres},
  {et~al\mbox{.}}} \bibinfo{year}{2019}\natexlab{}.
\newblock \showarticletitle{{X60: A Programmable Testbed for Wideband 60 GHz
  WLANs with Phased Arrays}}.
\newblock \bibinfo{journal}{\emph{Computer Communications}}
  \bibinfo{volume}{133} (\bibinfo{date}{Jan.} \bibinfo{year}{2019}), pp.
  \bibinfo{pages}{77--88}.
\newblock


\bibitem[\protect\citeauthoryear{{Saleh} and {Valenzuela}}{{Saleh} and
  {Valenzuela}}{1987}]%
        {saleh1987statistical}
\bibfield{author}{\bibinfo{person}{A.~A.~M. {Saleh}} {and} \bibinfo{person}{R.
  {Valenzuela}}.} \bibinfo{year}{1987}\natexlab{}.
\newblock \showarticletitle{A Statistical Model for Indoor Multipath
  Propagation}.
\newblock \bibinfo{journal}{\emph{IEEE Journal on Selected Areas in
  Communications}} \bibinfo{volume}{5}, \bibinfo{number}{2}
  (\bibinfo{date}{February} \bibinfo{year}{1987}), pp.
  \bibinfo{pages}{128--137}.
\newblock
\showISSN{1558-0008}


\bibitem[\protect\citeauthoryear{{Spencer}, {Peel}, {Swindlehurst}, and
  {Haardt}}{{Spencer} et~al\mbox{.}}{2004}]%
        {spencer2004introduction}
\bibfield{author}{\bibinfo{person}{Q.~H. {Spencer}}, \bibinfo{person}{C.~B.
  {Peel}}, \bibinfo{person}{A.~L. {Swindlehurst}}, {and} \bibinfo{person}{M.
  {Haardt}}.} \bibinfo{year}{2004}\natexlab{}.
\newblock \showarticletitle{{An Introduction to the Multi-User MIMO Downlink}}.
\newblock \bibinfo{journal}{\emph{IEEE Communications Magazine}}
  \bibinfo{volume}{42}, \bibinfo{number}{10} (\bibinfo{date}{Oct}
  \bibinfo{year}{2004}), pp. \bibinfo{pages}{60--67}.
\newblock
\showISSN{1558-1896}


\bibitem[\protect\citeauthoryear{Testolina, Lecci, Polese, Giordani, and
  Zorzi}{Testolina et~al\mbox{.}}{2020}]%
        {testolina2020scalable}
\bibfield{author}{\bibinfo{person}{P. Testolina}, \bibinfo{person}{M. Lecci},
  \bibinfo{person}{M. Polese}, \bibinfo{person}{M. Giordani}, {and}
  \bibinfo{person}{M. Zorzi}.} \bibinfo{year}{2020}\natexlab{}.
\newblock \showarticletitle{{Scalable and Accurate Modeling of the Millimeter
  Wave Channel}}. In \bibinfo{booktitle}{\emph{International Conference on
  Computing, Networking and Communications (ICNC)}}.
\newblock


\bibitem[\protect\citeauthoryear{{WBA and NGMN Alliance}}{{WBA and NGMN
  Alliance}}{2019}]%
        {ranConv}
\bibfield{author}{\bibinfo{person}{{WBA and NGMN Alliance}}.}
  \bibinfo{year}{2019}\natexlab{}.
\newblock \bibinfo{title}{{RAN Convergence Paper}}.
\newblock \bibinfo{howpublished}{version 1.0}.   (\bibinfo{date}{Aug.}
  \bibinfo{year}{2019}).
\newblock


\bibitem[\protect\citeauthoryear{Zhang, Polese, Mezzavilla, Rangan, and
  Zorzi}{Zhang et~al\mbox{.}}{2017}]%
        {zhang2017ns3}
\bibfield{author}{\bibinfo{person}{M. Zhang}, \bibinfo{person}{M. Polese},
  \bibinfo{person}{M. Mezzavilla}, \bibinfo{person}{S. Rangan}, {and}
  \bibinfo{person}{M. Zorzi}.} \bibinfo{year}{2017}\natexlab{}.
\newblock \showarticletitle{{ns-3 Implementation of the 3GPP MIMO Channel Model
  for Frequency Spectrum above 6 GHz}}. In
  \bibinfo{booktitle}{\emph{Proceedings of the Workshop on ns-3}}
  \emph{(\bibinfo{series}{WNS3 2017})}. \bibinfo{publisher}{ACM},
  \bibinfo{address}{Porto, Portugal}, pp. \bibinfo{pages}{71--78}.
\newblock
\showISBNx{9781450352192}


\bibitem[\protect\citeauthoryear{{Zhou}, {Cheng}, {Han}, {Fang}, {Fang}, {He},
  {Long}, and {Liu}}{{Zhou} et~al\mbox{.}}{2018}]%
        {zhou2018ieee}
\bibfield{author}{\bibinfo{person}{P. {Zhou}}, \bibinfo{person}{K. {Cheng}},
  \bibinfo{person}{X. {Han}}, \bibinfo{person}{X. {Fang}}, \bibinfo{person}{Y.
  {Fang}}, \bibinfo{person}{R. {He}}, \bibinfo{person}{Y. {Long}}, {and}
  \bibinfo{person}{Y. {Liu}}.} \bibinfo{year}{2018}\natexlab{}.
\newblock \showarticletitle{{IEEE 802.11ay-Based mmWave WLANs: Design
  Challenges and Solutions}}.
\newblock \bibinfo{journal}{\emph{IEEE Communications Surveys Tutorials}}
  \bibinfo{volume}{20}, \bibinfo{number}{3} (\bibinfo{date}{thirdquarter}
  \bibinfo{year}{2018}), pp. \bibinfo{pages}{1654--1681}.
\newblock
\showISSN{2373-745X}


\end{thebibliography}

\end{document}